\documentclass[amsmath,amssymb,noshowpacs,aip,pop,
preprint,
citeautoscript  
]{revtex4-1}

\usepackage{amsmath}
\usepackage{graphicx}
\usepackage[colorlinks=true, allcolors=blue]{hyperref}

\begin{document}

\title[Full $f$ and $\delta f$ GK particle simulations]{Full $f$ and $\delta f$ gyrokinetic particle simulations of Alfv\'en waves and energetic particle physics}
\author{Zhixin Lu, Guo Meng, Roman Hatzky, Matthias Hoelzl, Philipp Lauber}


\begin{abstract}
In this work, we focus on the development of the particle-in-cell scheme and the application to the studies of  Alfv\'en waves and energetic particle physics in tokamak plasmas. The $\delta f$ and full $f$ schemes are formulated on the same footing adopting mixed variables and the pullback scheme for electromagnetic problems. The TRIMEG-GKX code [Lu et al. J. Comput. Phys. 440 (2021) 110384] has been upgraded using cubic spline finite elements and full $f$ and $\delta f$ schemes.
The EP-driven TAE  has been simulated for the ITPA-TAE case featured by a small electron skin depth $\sim 1.18\times10^{-3}\;{\rm m}$, which is a challenging parameter regime of electromagnetic simulations, especially for the full $f$ model. The simulation results using the $\delta f$ scheme are in good agreement with previous work. Excellent performance of the mixed variable/pullback scheme has been observed for both full $f$ and $\delta f$ schemes.  Simulations with mixed full $f$ EPs and $\delta f$ electrons and thermal ions demonstrate the good features of this novel scheme in mitigating the noise level. The full $f$ scheme is a natural choice for EP physics studies which allows a large variation of EP profiles and distributions in velocity space, providing a powerful tool for kinetic studies using realistic experimental distributions related to intermittent and transient plasma activities.
\end{abstract}

\maketitle

\section{Introduction}
The gyrokinetic particle-in-cell (PIC) simulation provides a powerful tool for the studies of tokamak plasmas \cite{lee1983gyrokinetic}. 
For improving the simulation quality, especially for electromagnetic simulations, various schemes have been implemented such as the $p_\parallel$ formulae and the iterative scheme for solving Amp\'ere's law \cite{chen2007electromagnetic,hatzky2019reduction}, the noisy matrix \cite{mishchenko2005gyrokinetic} and the implicit scheme \cite{lu2021development,sturdevant2021verification}.  The noise reduction scheme has been summarized comprehensively recently \cite{hatzky2019reduction}, where various numerical applications have been studied using models with single ion species for linear physics in GYGLES, which demonstrates the excellent performance of the control variate method in noise reduction and the enhancement of simulation quality. 
The pullback scheme using mixed variables is implemented in ORB5 using the $\delta f$ scheme for the studies of EP driven TAEs \cite{mishchenko2019pullback}, demonstrating its capability in the MHD limit. In Table \ref{table:compare_fullf_deltaf}, we briefly summarized the discretization schemes (full $f$, control variate or $\delta f$) and the physics models (“symplectic ($v_\parallel$)” formula, “Hamiltionian ($p_\parallel$)” formula or mixed variables with pullback scheme) of some previous works. 


While most previous work has adopted the $\delta f$ scheme \cite{konies2018benchmark}, more effort has recently been spent on the full $f$ approach \cite{heikkinen2008full,chang2017fast,lu2021development}. The full $f$ method does not rely on the separation of the equilibrium and the perturbation, and thus provides a natural way to handle substantial changes of the profiles in the course of a simulation \cite{heikkinen2008full}. However, the full $f$ simulations are more expensive and require more strict noise reduction to make simulation studies of tokamak plasma feasible. 
Especially, the full $f$ model in the MHD limit is still a challenge. 
In this work, we focus on applications of the noise reduction schemes to the full $f$ and $\delta f$ simulations of Alfv\'en waves and energetic particle physics. By following the formulation from previous work \cite{hatzky2019reduction}, we implement the pullback scheme using mixed variables in the TRIMEG-(GKX) code \cite{lu2019development,lu2021development}. TRIMEG (TRIangular MEsh based Gyrokinetic) code was originally developed using the unstructured triangular meshes for the whole plasma volume simulations of the electrostatic ion temperature gradient mode \cite{lu2019development}, and later was extended to study Alfv\'en waves and energetic particle physics using the the full $f$ electromagnetic model \cite{lu2021development}. In the following, the full $f$ and $\delta f$ models are formulated and implemented on the same footing. 
Although the full $f$ scheme can be applied to all species in the TRIMEG-(GKX) code, a mixed scheme of full $f$ EPs and $\delta f$ thermal ions and electrons is proposed and applied in the present work. A common phenomenon in experiments is that the EP distribution in velocity space changes substantially, while the background is well described by the Maxwell distribution. With this novel mixed full $f$ and $\delta f$ scheme, the large EP profile variation and the arbitrary  distribution in velocity space can be treated in a natural way and the computational performance is improved compared to using a full $f$ scheme for all species. The mixed scheme for different species has been implemented in the TRIMEG-(GKX) code benefiting from its object-oriented programming and modular design of different species. 

The paper is organized as follows. In Sec. \ref{sec:model}, the equations for the discretization of full $f$ are derived with mixed variables and the pullback scheme adopted. In Sec. \ref{sec:numeric}, the normalized equations are given and the numerical methods are introduced with a rigorous filter derived. In Sec. \ref{sec:results}, the $\delta f$ and full $f$ simulations of energetic particle driven toroidicity induced Alfv\'en eigenmode are performed, demonstrating the features of the schemes and keys issues for the accurate description of the energetic particles in tokamak plasmas. 

\section{Physics models}
	\label{sec:model}
\subsection{Discretization of distribution function}
Following the formulation in the previous $\delta f$ work \cite{hatzky2019reduction,lanti2020orb5}, $N$ markers are used with a given distribution,
\begin{align}
    g(z,t)\approx \sum_{p=1}^{N} \frac{\delta[z_p-z_p(t)]}{J_z}\;\;,
\end{align}
where $z$ is the phase space coordinate, $\delta$ is the Dirac delta function, $J_z$ is the corresponding Jacobian and $z=({\bf R},v_\|,\mu\equiv v_\perp^2/(2B))$ is adopted in this work, ${\bf R}$ is the real space coordinate. For the full $f$ model, the total distribution of particles is represented by the markers,
\begin{align}
    f(z,t)=C_{\rm{g2f}} P_{\rm{tot}}(z,t)g(z,t)
    \approx C_{\rm{g2f}} \sum_{p=1}^{N} p_{p,\rm{tot}}(t) \frac{\delta[z_p-z_p(t)]}{J_z}\;\;,
\end{align}
where the constant $C_{\rm{g2f}}\equiv N_f/N_g$, $N_{f/g}$ is the number of particles/markers, and $g$ and $f$ indicate the markers and physical particles respectively. 
For each marker, 
\begin{align}
    p_{p,\rm{tot}}(t)=\frac{1}{C_{\rm{g2f}}}\frac{f(z_p,t)}{g(z_p,t)}=\text{const} \;\;,
\end{align}
for collisionless plasmas since 
\begin{align}
    \frac{{\rm d}g(z,t)}{{\rm d}t}=0 \;\;, \;\;
    \frac{{\rm d}f(z,t)}{{\rm d}t}=0 \;\;. 
\end{align}
The expression of $P_{\rm{tot}}(z,t)$ (and consequently, $p_{p,\rm{tot}}$) can be readily obtained 
\begin{align}
\label{eq:Ptot}
    P_{\rm{tot}}(z,t) = \frac{1}{C_{\rm{g2f}}}\frac{ f(z,t)}{g(z,t)}=\frac{n_f}{\langle n_f\rangle_V}\frac{\langle n_g \rangle_V}{n_g} \frac{f_v}{g_v} \;\;,
\end{align}
where  $n_f$ is the density profile and $f_v$ is the distribution in velocity space, namely, the particle distribution function $f=n_f({\bf R})f_v(v_\parallel,\mu)$, $\langle \ldots\rangle_V$ indicates the volume average. There are different choices of the marker distribution functions as discussed previously \cite{hatzky2019reduction,lanti2020orb5}. In this work, the markers are randomly distributed in toroidal direction and in the $(R,Z)$ plane but the distribution in velocity space is identical to that of the physical particles, which leads to 
\begin{align}
    P_{\rm{tot}}(z,t) = \frac{n_f}{\langle n_f\rangle_V}\frac{R}{R_0} \;\;.
\end{align}
The density and parallel current are readily obtained from the markers,
\begin{align}
    \{n,j_\|\}^{i_r,i_\theta,i_\phi} 
    &\equiv \int {\rm d} V \{n,j_\|\}({\bf{R}})N_{i_r}(r)N_{i_\theta}(\theta)N_{i_\phi}(\phi) \\
    &= C_{\rm{g2f}}\sum_{} p_{p,\rm{tot}} \{1,v_\|\} N_{i_r}(r_p)N_{i_\theta}(\theta_p)N_{i_\phi}(\phi_p)\;\;,
\end{align}
where ${\rm d}V=rR{\rm d}r\,{\rm d}\theta\, {\rm d}\phi$ for an ad-hoc equilibrium, $N_{i_r}$, $N_{i_\theta}$ and $N_{i_\phi}$ are basis functions and $i_r$, $i_\theta$ and $i_\phi$ indicate the indices in $r$, $\theta$ and $\phi$ directions. 

For the $\delta f$ model, the total distribution function is decomposed to the background and perturbed parts, $f(z,t)=f_0(z,t)+\delta f(z,t)$. The background part can be chosen as the time-independent one, i.e., $f_0(z,t)=f_0(z)$, and one typical choice is the Maxwellian distribution. The background and perturbed distribution functions are represented by the markers as follows,
\begin{align}
    f_0(z,t)=P(z,t)g(z,t)\approx \sum_{p=1}^{N} p_p(t) \frac{\delta[z_p-z_p(t)]}{J_z}\;\;, \\
    \delta f(z,t)=W(z,t)g(z,t)\approx \sum_{p=1}^{N} w_p(t) \frac{\delta[z_p-z_p(t)]}{J_z}\;\;,
\end{align}
where $p_{p}(t)=f_0(z_p,t)/g(z_p,t)$ and $w_{p}(t)=\delta f(z_p,t)/g(z_p,t)$ are time-varying variables. The evolution equations are readily obtained \cite{lanti2020orb5},
\begin{align}
    &\frac{\rm{d}}{{\rm d}t}w_p(t)  = - p_i(t) \frac{{\rm d}}{{\rm d}t} \ln f_0(z_p(t)) \;\;, \\
    &\frac{{\rm d}}{{\rm d}t}p_p(t) =   p_i(t) \frac{{\rm d}}{{\rm d}t} \ln f_0(z_p(t)) \;\;, \\
    &\frac{{\rm d}}{{\rm d}t} = \frac{\partial}{\partial t} + {\bf\dot R}\cdot\nabla + \dot v_\|\frac{\partial}{\partial v_\|} \;\;,
\end{align}
where $\dot \mu=0$ is used in the last equation. Generally, the guiding center's equation of motion can be decomposed to the equilibrium part corresponding to that in the equilibrium magnetic field, and the perturbed part due to the perturbed field 
\begin{eqnarray}
  {\bf\dot R} ={\bf\dot R}_0 +\delta {\bf\dot R} \;\;, \\
  \dot v_\| = \dot v_{\|,0}+\delta \dot{v} _\| \;\;.
\end{eqnarray}
For the equilibrium distribution function,
\begin{eqnarray}
 \left.\frac{{\rm d}}{{\rm d}t}\right|_0f_0=\left[\frac{\partial}{\partial t} + {\bf\dot R}_0\cdot\nabla + \dot v_{\|,0} \frac{\partial}{\partial v_\|}\right] f_0 =0\;\;,
\end{eqnarray}
and thus
\begin{eqnarray}
 \frac{{\rm d}}{{\rm d}t} f_0 = \left[\delta {\bf\dot R}\cdot\nabla + \delta \dot v_{\|} \frac{\partial}{\partial v_\|}\right]f_0 \;\;,
\end{eqnarray}
where $f_0$ is chosen as a steady state solution ($\partial f_0/\partial t=0$).
In this work, the Maxwell distribution is chosen ($f_0=f_{\rm{M}}$), 
\begin{eqnarray}
 f_{\rm{M}}=\frac{n_0}{(2T/m)^3\pi^{3/2}} \exp\left\{-\frac{mv_\|^2}{2T}-\frac{m\mu B}{T}\right\} \;\;,
\end{eqnarray}
and thus
\begin{eqnarray}
 \frac{{\rm d}}{{\rm d}t}\ln f_{\rm{M}} = \delta {\bf\dot R}\cdot\left[
 {\vec\kappa}_n + \left(\frac{mv_\|^2}{2T}+\frac{m\mu B}{T}-\frac{3}{2}\right)\vec\kappa_T
 -\frac{m\mu B}{T}\vec\kappa_B
 \right]
 - \delta \dot v_{\|} \frac{mv_\|}{T} \;\;, \nonumber\\
\end{eqnarray}
where $\vec\kappa_{n,T,B}\equiv \nabla\ln \{n,T,B\}$. 
Note for the Maxwell distribution, without considering the neoclassical physics, the following approximation has been made in the traditional $\delta f$ scheme,
\begin{eqnarray}
 \left.\frac{{\rm d}}{{\rm d}t}\right|_0f_{\rm{M}} \approx 0\;\;.
\end{eqnarray}

\begin{table}
\centering
\begin{tabular}{l| c | c | r}
        & Full $f$ & control variate  & traditional $\delta f$ \\\hline
 $v_\|$     & \citenum{lu2021development}  &  & \citenum{sturdevant2021verification} \\ \hline
 $p_\|$     &  & \citenum{hatzky2019reduction} & \citenum{chen2007electromagnetic,bottino2011global} \\ \hline
 MV w/ PB  & \citenum{hager2022electromagnetic}, this work & \citenum{hatzky2019reduction} & \citenum{mishchenko2019pullback}, this work \\ \hline
\end{tabular}
\caption{\label{table:compare_fullf_deltaf} Various works using full $f$ and $\delta f$ schemes. MV and PB indicate Mixed Variable and pullback schemes, respectively. }
\end{table}

\subsection{Physics equations using mixed variables}
The mixed variable is defined as follows.
The parallel component of the scalar potential is decomposed to the symplectic part and the Hamiltonian part,
\begin{equation}
    \delta A_\| =\delta A_\|^{\rm{s}} + \delta A_\|^{\rm{h}} \;\;,
\end{equation}
where the symplectic part is chosen to satisfy 
\begin{equation}
    \partial_t\delta A_\|^{\rm{s}}+\partial_\|\delta\phi = 0\;\;.
\end{equation}
The parallel velocity coordinate of the guiding center is defined as 
\begin{equation}
    u_\|=v_\|+\frac{q_s}{m_s}\langle\delta A_\|^{\rm{h}}\rangle\;\;,
\end{equation}
where $q_s$ and $m_s$ are the charge and mass of species $s$, respectively, the subscript $s$ represents the different particle species, and $\langle\ldots\rangle$ indicates the gyro average .

The guiding center's equations of motion are consistent with previous work \cite{mishchenko2019pullback,hatzky2019reduction,lanti2020orb5}, 
\begin{align*}
  &\frac{{\rm d}{\bf R_0}}{{\rm d}t} = v_\| {\bf b}^* + \frac{m\mu}{qB^*} {\bf b}\times\nabla B \;\;, 
  \\
  &\frac{{\rm d}u_{\|,0}}{{\rm d}t} = -\mu {\bf b}^*\cdot \nabla B \;\;,
  \\
  &\frac{{\rm d}\delta{\bf R}}{{\rm d}t} = \frac{{\bf b}}{B^*}\times \nabla \langle \delta\phi -v_\| \delta A_\|\rangle \;\;, 
  \\
  &\frac{{\rm d}\delta u_\|}{{\rm d}t} =  -\frac{q_s}{m_s} \left({\bf b}^*\cdot\nabla\langle\delta\phi-v_\|\delta A^{\rm{h}}_\|\rangle +\partial_t\langle\delta A_\|^{\rm{s}}\rangle \right) -\frac{\mu}{B^*}{\bf b}\times\nabla B\cdot\nabla\langle\delta A_\|^{\rm{s}}\rangle \;\;,
\end{align*}
where $v_\|$ is adopted on the right hand side, and thus, the term $-(q_s/m_s)\langle\delta A_\|^{\rm{h}}\rangle{\bf b}^*$ in ${\rm d}{\bf R}/{\rm d}t$ is taken into account in ${\rm d}{\bf R}_0/{\rm d}t$. 

The quasi-neutrality equation is 
\begin{equation}
    -\nabla\cdot\left( \sum_s\frac{q_s n_{0s}}{B\omega_{cs}} \nabla_\perp\delta\phi \right) = \sum_s q_s \delta n_{s,v} \;\;,
\end{equation}
where $\delta n_s$ is calculated using $\delta f({\bf R},v_\|,\mu)$ (indicated as $\delta f_{s,v}$), namely, $\delta n_s({\mathbf{x}})=\int d^6 z\delta f_s\delta({\mathbf{R+\rho-x}})$, $\omega_{cs}$ is the cyclotron frequency of species `$s$' and in this work, we ignore the perturbed electron polarization density on the left hand side. 
When the $\delta f$ scheme is adopted, $\delta f_{s,v}$ is obtained from $\delta f_{s,u}$ as follows with the linear approximation of the pullback scheme,
\begin{eqnarray}
    & \delta f_{v} = \delta f_{u} +  \frac{q_s\left\langle\delta A^{\rm{h}}_{\|} \right\rangle}{m_s}\frac{\partial f_{0s}}{\partial v_\|}
    \xrightarrow[f_{0s}=f_{\rm{M}}]{\text{Maxwellian}}
     \delta f_{u} -  \frac{ m_sv_\|}{T_s}  \frac{q_s\left\langle\delta A^{\rm{h}}_{\|} \right\rangle}{m_s} f_{0s} \;\;,
\end{eqnarray}
which is obtained from the more general form,
\begin{eqnarray}
    & f_{v} (v_\|) = f_{u} (v_\|+\frac{q_s}{m_s}\langle\delta A_\|^h\rangle)\;\;.
\end{eqnarray}

Amp\'ere's law in $v_\|$ space is given by 
\begin{equation}
    -\nabla^2_\perp\delta A_\| = \mu_0 \delta j_{\|,v} \;\;,
\end{equation}
where $\delta j_{\|,v}({\mathbf{x}})=\int d^6 z\delta f_{v,s}\delta({\mathbf{R+\rho-x}})v_\|$.

For the $\delta f$ model, using the mixed variables and assuming Maxwell distribution, we have 
\begin{eqnarray}
    \delta j_{\|,v}
    &\equiv&\sum_s q_s\int \mathrm{d}z^6\delta f_s(v_\|)\delta({\mathbf{R+\rho-x}}) v_\| \nonumber \\
    &=&\sum_s q_s\int \mathrm{d}z^6 \left[\delta f_s(u_\|) -\frac{v_\| q_s\langle \delta A_\|^{\rm{h}}\rangle}{T_s}\right] \delta({\mathbf{R+\rho-x}})  v_\| \;\;.
\end{eqnarray}
Then we can write Amp\'ere's law as
\begin{eqnarray}
    -\nabla^2_\perp\delta A_{\|}^{\rm{h}}
    +\sum_s\mu_0\frac{e_s^2}{T_s}\int \mathrm{d}z^6 v_\|^2 f_0 \langle \delta A_{\|}^{\rm{h}} \rangle \delta({\mathbf{R+\rho-x}}) \nonumber\\
    =\nabla^2_\perp\delta A_{\|}^{\rm{s}} 
    + \mu_0\sum_s q_s  \int \mathrm{d}z^6  v_\| \delta f_s(u_\|)\delta({\mathbf{R+\rho-x}})   \;\;.
\end{eqnarray}
The integral on the left-hand side can be obtained analytically, yielding
\begin{eqnarray}
\label{eq:ampere_mv_deltaf}
    -\nabla^2_\perp\delta A_{\|}^{\rm{h}}
    +\sum_s\frac{1}{d_{s}^2}\overline{\langle\delta A_{\|}^{\rm{h}}\rangle  }
    &=&\nabla^2_\perp\delta A_{\|}^{\rm{s}} 
     \nonumber \\
     &+&\mu_0\sum_s q_s  \int \mathrm{d}z^6  v_\| \delta f_s(u_\|)  \delta({\mathbf{R+\rho-x}}) \;\;, \\
    \overline{\langle\delta A_{\|}^{\rm{h}}\rangle  }
    &\equiv&\frac{2}{n_{s0}v_{ts}}\int \mathrm{d}z^6 v_\|^2 f_0 \langle \delta A_{\|}^{\rm{h}} \rangle \delta({\mathbf{R+\rho-x}}) \,\,,
\end{eqnarray}
where $v_{\rm{t}s}=\sqrt{2T_s/m_s}$, $d_{s}$ is the skin depth of species `$s$' defined as $d_{s}^2=c^2/\omega_{p,s}^2=m_s/(\mu_0q_s^2n_{0s})$. 

For the full $f$ model, the perturbed current is represented by full $f$,
\begin{eqnarray}
    \delta j_{\|,v}=\sum_s q_s\int \mathrm{d}v^3 v_\| f_s =  \sum_sq_s\int \mathrm{d}z^6 \left[ u_\|-\frac{q_s}{m_s}\langle\delta A_\|^{\rm{h}}\rangle\right] f_s \delta({\mathbf{R+\rho-x}})\;\;. \nonumber\\
\end{eqnarray} 
Amp\'ere's law yields
\begin{eqnarray}
    -\nabla^2_\perp\delta A_{\|}^{\rm{h}}
    +\sum_s\mu_0\frac{e_s^2}{m_s}\int \mathrm{d}z^6 f_s\langle \delta A_{\|}^{\rm{h}} \rangle\delta({\mathbf{R+\rho-x}})  \nonumber \\
    =\nabla^2_\perp\delta A_{\|}^{\rm{s}} + \mu_0\sum_s q_s  \int \mathrm{d}z^6  u_\| f_s \delta({\mathbf{R+\rho-x}}) \;\;.
\end{eqnarray}
The corresponding analytical limit gives the similar form of Eq. \ref{eq:ampere_mv_deltaf} except the replacement of $\delta f(u_\|)$ with $f(u_\|)$ and the definition of $\overline{\langle\delta A_{\|}^{\rm{h}}\rangle  }$,
\begin{eqnarray}
    -\nabla^2_\perp\delta A_{\|}^{\rm{h}}
    +\sum_s\frac{1}{d_{s}^2}\overline{\langle\delta A_{\|}^{\rm{h}}\rangle  }& =&
    \nabla^2_\perp\delta A_{\|}^{\rm{s}} 
     \nonumber \\
    &+&\mu_0\sum_s q_s  \int \mathrm{d}z^6  v_\| f_s(u_\|)  \delta({\mathbf{R+\rho-x}})  \;\;, \\
    \overline{\langle\delta A_{\|}^{\rm{h}}\rangle  }
    &\equiv&
    \frac{1}{n_{s0}}\int \mathrm{d}z^6  f_0 \langle \delta A_{\|}^{\rm{h}} \rangle \delta({\mathbf{R+\rho-x}}) 
\end{eqnarray}

Using the iterative scheme, the asymptotic solution is expressed as follows,
\begin{equation}
    \delta A^{\rm{h}}_{\|}=\sum_{p=0}^\infty\delta A^{\rm{h}}_{\|,p}\;\;,
\end{equation}
where $\epsilon=|\delta A^{\rm{h}}_{\|,p+1}/\delta A^{\rm{h}}_{\|,p}|\ll1$. 
Amp\'ere's law is solved order by order,
\begin{eqnarray}
\label{eq:ampere_h0}
    \left(\nabla^2_\perp-\sum_s\frac{1}{d_{s}^2}\right)\delta A_{\|,0}^{\rm{h}} 
    = -\nabla^2_\perp\delta A_{\|}^{\rm{s}} - \mu_0 \delta j_{\|,p} \;\;, \\
    \label{eq:ampere_iterative}
    \left(\nabla^2_\perp-\sum_s\frac{1}{d_{s}^2}\right)\delta A_{\|,p}^{\rm{h}} 
    =-\sum_s\frac{1}{d_{s}^2}\delta A^{\rm{h}}_{\|,p-1} 
    + \sum_s\frac{1}{d_{s}^2} \overline{\langle\delta A_{\|,p-1}^{\rm{h}}\rangle}\;\;, \\
    \overline{\langle\delta A_{\|,p-1}^{\rm{h}}\rangle}
    =\frac{2}{n_0 v_{\rm{t}s}^2}\int \mathrm{d}z^6 v_\|^2 f_0  \langle\delta A^{\rm{h}}_{\|,p-1} \rangle\delta({\mathbf{R+\rho-x}})  \;\;,\text{ for $\delta f$ model} \\
\label{eq:A2ndavg_fullf}
    \overline{\langle\delta A_{\|,p-1}^{\rm{h}}\rangle}
    =\frac{1}{n_0}\int \mathrm{d}z^6 f  \langle\delta A^{\rm{h}}_{\|,p-1} \rangle \delta({\mathbf{R+\rho-x}}) \;\;,\text{ for full $f$ model}
\end{eqnarray}
where  $p=1,2,3,\ldots$ and since $2/(n_0 v_{\rm{t}s}^2)\int \mathrm{d}v^3 v_\|^2 f_0=1$ and $(1/n_0)\int \mathrm{d}v^3 v_\|^2 f_0=1$ for the Maxwell distribution in the analytical limit, good convergence of the iterative solver is expected.

\subsection{Pullback scheme for mitigating the cancellation problem}
More detailed description of the pullback scheme can be found in the previous work \cite{mishchenko2019pullback}. As a brief review, the equations for the $\delta f$ are listed as follows.
\begin{align}
\label{eq:pullback_A}
    & \delta A^{\rm{s}}_{\|,\rm{new}} = \delta A^{\rm{s}}_{\|,\rm{old}} + \delta A^{\rm{h}}_{\|,\rm{old}}  \;\;, \\
\label{eq:pullback_v}
    & u_{\|,\rm{new}} = u_{\|,\rm{old}} - \frac{q_s}{m_s} \left\langle\delta A^{\rm{h}}_{\|,\rm{old}} \right\rangle  \;\;, \\
\label{eq:pullback_df}
    & \delta f_{\rm{new}} = \delta f_{\rm{old}} +  \frac{q_s\left\langle\delta A^{\rm{h}}_{\|,\rm{old}} \right\rangle}{m_s}\frac{\partial f_{0s}}{\partial v_\|}
    \xrightarrow[f_{0s}=f_{M}]{\text{Maxwellian}}
     \delta f_{\rm{old}} -  \frac{ 2v_\|}{v_{\rm{t}s}^2}  \frac{q_s\left\langle\delta A^{\rm{h}}_{\|,\rm{old}} \right\rangle}{m_s} f_{0s} \;\;,
\end{align}
where Eq. \ref{eq:pullback_df} is the linearized equation for $\delta f$ pullback, which is from the general equation of the transformation for the distribution function 
\begin{align}
    f_{\rm{old}} (u_{\| \rm{old}}) = f_{\rm{new}} (u_{\| \rm{new}} =u_{\| \rm{old}}- \frac{q_s}{m_s} \left\langle\delta A^{\rm{h}}_{\|,\rm{old}} \right\rangle ) \;\;. 
\end{align}
For the full $f$ scheme, only Eqs. \ref{eq:pullback_A} and \ref{eq:pullback_v} are needed. 

\subsection{Kinetic equilibrium in constant of motion coordinates}
While the local Maxwellian distribution is widely used in gyrokinetic simulation, its application in the full $f$ scheme brings in marker relaxation and a consequent lower growth rate for the EP driven TAE problem. The shifted toroidal canonical momentum has been adopted in the $\delta f$ particle code ORB5 for the turbulence studies \cite{angelino2006definition}. In this work, we apply this scheme also in the full $f$ model. To construct the marker distribution in the constant of motion space, we take the shifted toroidal canonical momentum 
\begin{eqnarray}
\label{eq:psishift}
    \psi_{\rm{can}}=\psi+\frac{m_s F}{q_s B} v_\| - \text{sign}(v_\|)\sqrt{2(E-\mu B_0)}\frac{m F}{qB_0}H(E-\mu B_0)\;\;,
\end{eqnarray}
where the last term gives the size of the finite orbit width. For energetic particles, among the right hand side terms in Eq.  \ref{eq:psishift}, the second term can be of the same magnitude of the first term and it is a natural choice to bring in the shit (the last term) so that $\psi_{can}$ is close to the orbit center. 
The distribution is specified as
\begin{eqnarray}
\label{eq:fcanEP}
    f_{\rm{can}}(\psi_{\rm{can}},E,\mu)=n(\psi_{\rm{can}})\exp\left\{-\frac{mE}{T(\psi_{\rm{can}})}\right\}\;\;,
\end{eqnarray}
where the variation along $\mu$ direction is eliminated for the sake of simplicity. Correspondingly, when loading markers, $n(r_{\rm{can}})$ replaces $n(r)$ in Eq. \ref{eq:Ptot}, where $r_{\rm{can}}^2\equiv(\psi_{\rm{can}}-\psi_{\rm{axis}})/(\psi_{edge}-\psi_{\rm{axis}})$, $r_{\rm{can}}\ge0$ for $(\psi_{\rm{can}}-\psi_{\rm{axis}})/(\psi_{\rm{edge}}-\psi_{\rm{axis}})\ge0$ and $r_{\rm{can}}<0$ for $(\psi_{\rm{can}}-\psi_{\rm{axis}})/(\psi_{\rm{edge}}-\psi_{\rm{axis}})<0$. 

\section{Numerical schemes}
\label{sec:numeric}
\subsection{Normalized equations}
The normalization of the variables in the TRIMEG-GKX code is introduced in this section. The length unit is $R_{\rm{N}}=1$ m. The particle mass is normalized to $m_{\rm{N}}$ and $m_{\rm{N}}=m_{\rm{e}}$. 
The velocity unit is 
\begin{align*}
    v_{\rm{N}}\equiv \sqrt{2T_{\rm{N}}/m_{\rm{N}}} \;\;,
\end{align*}
where $T_{\rm{N}}$ and $m_{\rm{N}}$ are the temperature and mass unit for normalization. 

Temperature is normalized to $T_{\rm{N}}=m_{\rm{N}}v_{\rm{N}}^2/2$, namely, 
\begin{eqnarray}
  T_s=\bar{T}_s T_{\rm{N}} = \frac{1}{2} \bar{T}_s m_{\rm{N}} v_{\rm{N}}^2 \;\;.
\end{eqnarray}
Note another way (not adopted in this work) is to normalize $T$ to $2T_{\rm{N}}$, namely,
    $T_s=\bar{T}_s m_{\rm{N}} v_{\rm{N}}^2 
    =2 \bar{T}_s T_{\rm{N}}$.
In addition, $\mu$ is normalized to $v_{\rm{N}}^2/B_{\rm{ref}}$, 
\begin{align*}
    \mu\equiv\frac{ v_\perp^2}{2B}=\bar{\mu} \frac{v_{\rm{N}}^2}{B_{\rm{ref}}},
\end{align*}
where $B_{\rm{ref}}=1\;{\rm T}$. 

The Maxwell distribution is 
\begin{eqnarray}
  f_{\rm{M}}
  = \frac{1}{v_{\mathrm t}^3\pi^{3/2}} e^{-\frac{mv_\|^2+2\mu B}{2T}}
  = \frac{1}{v_{\mathrm t}^3\pi^{3/2}} \exp \left\{-\frac{\bar{m}\bar{v}^2_\|}{\bar{T}} -2\frac{\bar{m}\bar{\mu}_\|}{\bar{T}} \frac{B}{B_{\rm{ref}}} \right\} 
  \;\;,
\end{eqnarray}
and correspondingly,
\begin{eqnarray}
 \frac{{\rm d}}{{\rm d}\bar{t}}\ln f_{\rm{M}}& = &\frac{{\rm d}\delta {\bf\bar R}}{{\rm d}\bar{t}}\cdot\left[
 {\vec\kappa}_n + 
 \left(\frac{\bar m \bar v_\|^2}{\bar T}+\frac{2\bar m \bar\mu \bar B}{T}-\frac{3}{2}\right)\vec\kappa_T
 -2\frac{\bar m\bar\mu \bar B}{\bar T}\vec\kappa_B
 \right] \nonumber\\
 &-& \frac{2\bar m\bar v_\|}{\bar T}  \frac{{\rm d}\delta \bar v_\|}{{\rm d}\bar{t}}\;\;.
\end{eqnarray}
The markers are loaded with the same distribution of physical particles in velocity space but uniformly in the poloidal plane and in the toroidal direction. In $v_\|$ direction, a random number generator is used to produce numbers with normal distribution $f(x)=1/(\sigma\sqrt{2\pi})\exp\{-[(x-x_0)/\sigma]^2/2\}$, where $x_0$ and $\sigma$ are chosen as $0$ and $\sqrt{\bar{T}/(2\bar{m})}$ respectively. In $\mu$ direction, the uniformly distributed random numbers~$x$ are generated and shifted according to $\mu=-\ln(x)\bar{T}B_{\rm{ref}}/(2\bar{m}B)$. 

The normalized guiding center's equations of motion are
\begin{eqnarray}
  \frac{{\rm d}{\bf R_0}}{{\rm d}t} 
  &=& v_\| {\bf b}^* + \frac{\Bar{m}_s}{\Bar{q}_s}\Bar{\rho}_{\rm{N}}\frac{B_{\rm{ref}}}{B^2B^*} \mu{\bf B}\times\nabla B \;\;, 
  \\
  \frac{{\rm d}u_{\|,0}}{{\rm d}t} 
  &=& -\mu {\bf b}^*\cdot \nabla B \;\;,
  \\
  \frac{{\rm d}\delta{\bf R}}{{\rm d}t} 
  &=& \Bar{\rho}_{\rm{N}}\frac{B_{\rm{ref}}}{B^*}{\bf b}\times \nabla \langle \delta\phi -v_\| \delta A_\|\rangle \;\;, 
  \\
  \frac{{\rm d}\delta u_\|}{{\rm d}t} 
  &=&  -\frac{\Bar{q}_s}{\Bar{m}_s} \left({\bf b}^*\cdot\nabla\langle\delta\phi-v_\|\delta A^{\rm{h}}_\|\rangle +\partial_t\langle\delta A_\|^{\rm{s}}\rangle \right)  \nonumber\\ &-&\rho_{\rm{N}}\frac{B_{\rm{ref}}}{B^*}\mu{\bf b}\times\nabla B\cdot\nabla\langle\delta A_\|^{\rm{s}}\rangle \;\;,
\end{eqnarray}
where $\delta \phi$ and $\delta A_\|$ are normalized to $m_{\rm{N}} v_{\rm{N}}^2/e$ and $m_{\rm{N}} v_{\rm{N}}/e$ respectively. 

The normalized quasi-neutrality equation is,
\begin{eqnarray}
    \nabla_\perp \sum n_{0s} \bar{M}_s (B_{\rm{axis}}/B)^2\nabla_\perp\delta \bar{\phi}=C_{\mathrm P}\sum(-\bar{e}_s)\delta n_s\;\;,
\end{eqnarray}
where $C_{\mathrm P}=1/\rho_{\rm{N}}^2$.

For Amp\'ere's law, the original normalized equation $\nabla^2_\perp\delta A =C_{\mathrm A} \delta J_{\|,\rm{ph}}$, where $C_{\mathrm A} =\beta_{ref}/\rho_N^2$, is solved using mixed variables and the iterative scheme (corresponding to Eqs. \ref{eq:ampere_h0}--\ref{eq:A2ndavg_fullf}), 
\begin{eqnarray}
    &&\left(\nabla^2_\perp-\sum_s\frac{\bar{q}_s^2}{M_s}C_{\mathrm A}\right)\delta A_{\|,0}^{\rm{h}} 
    = -\nabla^2_\perp \delta A_{\|,0}^{\rm{s}} - C_{\mathrm A} \delta J_{\|} \;\;, 
    \\
    &&\left(\nabla^2_\perp-\sum_s\frac{\bar{q}_s^2}{M_s}C_{\mathrm A}\right)\delta A_{\|,0}^{\rm{h}} 
    = -\sum_s\frac{\bar{q}_s^2}{M_s}C_{\mathrm A}\delta A_{\|,p-1}^{\rm{h}} 
    + \bar{G} \delta A_{\|,p-1}^{\rm{h}} 
    \\
    &&\bar{G}\delta A_{\|,p-1}^{\rm{h}} =C_{\mathrm A}\frac{N_{0s} \bar{q}_s^2}{\bar{T}_s}\sum_p 2v_{\parallel,p}^2 
    \int dz^6 w_p\delta({\mathbf{R+\rho-x}})\langle\delta A_{\|,p-1}^{\rm{h}}\rangle   {\text{ for $\delta f$}}\;\;,
    \\
    &&\bar{G}\delta A_{\|,p-1}^{\rm{h}} =C_{\mathrm A}\frac{N_{0s} \bar{q}_s^2}{M_s}\sum_p 
    \int dz^6 p_{p,tot}\delta({\mathbf{R+\rho-x}}) 
    \langle\delta A_{\|,p-1}^{\rm{h}}\rangle   {\text{ for full $f$}} \;\;. 
\end{eqnarray}

The normalized equations for the pullback treatment are as follows,
\begin{align}
    & \delta A^{\rm{s}}_{\|,\rm{new}} = \delta A^{\rm{s}}_{\|,\rm{old}} + \delta A^{\rm{h}}_{\|,\rm{old}}  \;\;, \\
    & u_{\|,\rm{new}} = u_{\|,\rm{old}} - \frac{\bar{q}}{\bar{m}_s} \left\langle\delta A^{\rm{h}}_{\|,\rm{old}} \right\rangle  \;\;, \\
    & \delta f_{\rm{new}} = \delta f_{\rm{old}} +  \frac{\Bar{q}_s\left\langle\delta A^{\rm{h}}_{\|,\rm{old}} \right\rangle}{\bar{m}_s}\frac{\partial f_{0s}}{\partial \bar{v}_\|} \\
    \label{eq:pullback_df_norm}
    &
    \xrightarrow[f_{0s}=f_{\rm{M}}]{\text{Maxwellian}}
     \delta f_{\rm{old}} -  \frac{ 2 \bar{q}_s}{\bar{T}_s} \bar{v}_\| \left\langle\delta A^{\rm{h}}_{\|,\rm{old}} \right\rangle f_{0s} \;\;,
\end{align}
where the factor $2$ is from the normalization of $T$ to $T_{\rm{N}}=m_{\rm{N}}v_{\rm{N}}^2/2$, and Eq. \ref{eq:pullback_df_norm} is the linearized pullback scheme for the $\delta f$ model implemented in our work. The studies using the nonlinear pullback scheme is out of the scope of this work and will be addressed in future. 

\subsection{Finite element method}
The three dimensional solver is developed in this work with the finite element method adopted in the radial, poloidal and toroidal directions.  
Periodic boundary condition is adopted in poloidal and toroidal directions. In radial direction, the Dirichlet boundary condition with zero value of the function is implemented. 
The grids size is $(N_r,N_\theta,N_\phi)$ and $(N_{r,\rm{FEM}},N_{\theta,\rm{FEM}},N_{\phi,\rm{FEM}})$ basis functions are adopted to represent functions in the simulation domain, where $N_{r,\rm{FEM}}=N_r+\Delta N$, $N_{\theta,\rm{FEM}}=N_\theta$, $N_{\phi,\rm{FEM}}=N_\phi$, which are consistent with the boundary conditions, where $\Delta N=2$ since cubic splines are adopted. 
In poloidal and toroidal directions, the cubic finite element basis functions $N(x)$ are as follows
\begin{equation}
  N_{\rm{cubic}}(x) =
    \begin{cases}
       4/3+2x+x^2+x^3/6 \;\;, & \text{if $x\in[-2,-1)$}\\
       2/3-x^2-x^3/2 \;\;,    & \text{if $x\in[-1,0) $}\\
       2/3-x^2+x^3/2 \;\;,    & \text{if $x\in[0,1) $} \\
       4/3-2x+x^2-x^3/6 \;\;. & \text{if $x\in[1,2) $}
    \end{cases}       
\end{equation}
Along $\theta$ and $\phi$, the $i$th basis function is $N_i=N_{\rm{cubic}}(x+1-i)$. 
In radial direction, $N_i$ is the same as those in poloidal/toroidal directions as $i\ge4$ or $i\le N_{r,\rm{FEM}}-3$. The first basis function is 
\begin{equation}
  N_{\rm{cubic}}(x) =
    \begin{cases}
       0 \;\;, & \text{if $x\in[-2,1)$}\\
       -x^3+6x^2-12x+8 \;\;. & \text{if $x\in[1,2) $}
    \end{cases}       
\end{equation}
The second basis function is
\begin{equation}
  N_{\rm{cubic}}(x) =
    \begin{cases}
       0 \;\;, & \text{if $x\in[-2,0)$}\\
       7x^3/6-3x^2+2x \;\;,    & \text{if $x\in[0,1) $} \\
       4/3-2x+x^2-x^3/6 \;\;. & \text{if $x\in[1,2) $}
    \end{cases}       
\end{equation}
The third basis function is
\begin{equation}
  N_{\rm{cubic}}(x) =
    \begin{cases}
       0 \;\;, & \text{if $x\in[-2,-1)$}\\
       -x^3/3-x^2+2/3 \;\;,    & \text{if $x\in[-1,0) $}\\
       x^3/2-x^2+2/3 \;\;,    & \text{if $x\in[0,1) $} \\
       -x^3/6+x^2-2x+4/3 \;\;. & \text{if $x\in[1,2) $}
    \end{cases}       
\end{equation}
The last three basis functions are symmetric mapping of the first three basis functions with respect to the middle point of the simulation domain.  All radial basis functions are constructed according to  $N_i=N_{\rm{cubic}}(x+1-i)$, where $i\in[1,N_{r,\rm{FEM}}]$. 

\subsection{Weak form of field equations}
For a partial differential equation,  
\begin{eqnarray}
  L(r,\theta,\phi)y(r,\theta,\phi)=b(r,\theta,\phi) \;\;,
\end{eqnarray}
where $L$ is a linear differential operator, the weak form can be written as
\begin{eqnarray}
    \int {\rm d} r\, {\rm d}\theta\, {\rm d}\phi S(r,\theta,\phi) N_{i_r}N_{i_\theta}N_{i_\phi} 
    L(r,\theta,\phi)y(r,\theta,\phi)
    \nonumber \\
     =\int {\rm d} r\, {\rm d}\theta\, {\rm d}\phi S(r,\theta,\phi) N_{i_r}N_{i_\theta}N_{i_\phi}
    b(r,\theta,\phi) \;\;,
\end{eqnarray}
where $S(r,\theta,\phi)$ is a function and $S=1$ is chosen in this work. 
The weak form of the quasi-neutrality equation, Amp\'ere's law, the iterative equation, Ohm's law are
\begin{eqnarray}
    \bar{\bar{M}}_{P,L,ii',jj',kk'} \cdot\delta\phi_{i'j'k'}
    &=& C_{\mathrm P} \delta N^{i,j,k} \;\;,  \\
    \bar{\bar{M}}_{\mathrm{A},L,ii',jj',kk'} \cdot\delta A^{\rm{h}}_{i'j'k'} 
    &=& \bar{\bar{M}}_{\mathrm{A},R,ii',jj',kk'} \delta A^{\rm{s}}_{i'j'k'}+ C_{\mathrm A} \delta J^{i,j,k} \;\;,  \\
    \bar{\bar{M}}_{\mathrm{it},L,ii',jj',kk'} \cdot\delta A^{h}_{i'j'k',[p+1]} 
    &=& \bar{\bar{M}}_{\mathrm{it},R,ii',jj',kk'} \cdot\delta A^{h}_{i'j'k',[p]} 
    +\langle\delta A^{h,i'j'k',[p]} \rangle \;\;, \\
    \bar{\bar{M}}_{\mathrm{Ohm},L,ii',jj',kk'} \cdot\delta A^{\rm{s}}_{i'j'k'} 
    &=& 
    \bar{\bar{M}}_{\mathrm{Ohm},R,ii',jj',kk'} \cdot \delta\phi_{i'j'k'} 
\end{eqnarray}

\subsection{Fourier filter}
\subsubsection{Filter for moments (particle-in-Fourier)}
For moment variables ($\delta n$ and $\delta j_\parallel$), the Fourier components are calculated from markers first as follows,
\begin{align}
    \{\delta n_{n,m},\delta j_{\|,n,m} \}
    =C_{\rm{p2g}} \sum_p \{1,v_\|\} w_p e^{-\mathrm{i}n\phi_p-\mathrm{i} m\theta_p}\;\;,
\end{align}
where $n$ and $m$ are the toroidal and poloidal mode numbers of the filter and $C_{\rm{p2g}}$ is the converting factor from marker to grid variables.
The corresponding spline coefficients are readily obtained,
\begin{align}
    \{ \delta n,\delta j_{\|} \}_{i_r,i_\theta,i_\phi} =  \{\delta n,\delta j_\| \}_{n,m} T_{n,i_\phi} T_{m,i_\theta}\;\;,
\end{align}
where 
\begin{align}
    T_{n,i} = \int_{\phi_{\rm min}}^{\phi_{\rm max}} \mathrm{d}\phi e^{{\rm i}n\phi}N_i(\phi) 
    =\Delta\phi \int_{-2}^{2} {\mathrm{d}}x e^{{\rm i}n_{\rm{eff}}x}N_i(x)  \;\;,\\
    T_{m,i} = \int_{\theta_{\rm min}}^{\theta_{\rm max}} \mathrm{d}\theta e^{{\rm i}m\theta}N_i(\phi) 
     =\Delta\theta \int_{-2}^{2} {\rm d}x e^{{\rm i}m_{\rm{eff}}x}N_i(x) 
     \;\;,
\end{align}
where $n_{\rm{eff}}=n\Delta \phi$, $m_{\rm{eff}}=n\Delta \theta$, $\Delta \phi$ and $\Delta\theta$ are the grid size in toroidal and poloidal directions.
The analytic results are used for the construction,
\begin{equation}
    \int_{-2}^2 {\rm d}x N(x) e^{-{\rm i}nx} = \frac{6+2\cos(2n)-8\cos(n)}{n^4}  \;\;.
\end{equation}

\subsubsection{Filter for fields}
For field variables such as $\delta\phi$ and $\delta A_\|$, the physical values are expressed using the spline coefficients and the Fourier filter is applied to the physical values. 
The spline coefficients of a variable $\delta\phi$ are filtered in $\phi$ and $\theta$ directions as follows,
\begin{align}
    &\delta\Bar{\phi}_{i_\phi} = 2 \text{Re}\{ M_{i_\phi j_\phi}^{-1} [T_{n,i_\phi} T_{-n,j_\phi} \delta\phi_{j_\phi}]/\phi_{\rm{wid}} \} \;\;, \\
    &\delta\Bar{\phi}_{i_\theta} = 2 \text{Re}\{ M_{{i_\theta} j_\theta}^{-1} [T_{m,{i_\theta}} T_{-m,{j_\theta}} \delta\phi_{j_\theta}]/\theta_{\rm{wid}} \} \;\;,
\end{align}
where $\phi_{\rm{wid}}$ and $\theta_{\rm{wid}}$ are the width of the simulation domain in toridal and poloidal directions, respectively, $\phi_{\rm{wid}}=2\pi$ for full torus simulations, $\theta_{\rm{wid}}=2\pi$, and $M_{i,j}$ is the mass matrix.

\subsection{Parallelization and the application of shared memory MPI to field implementation}
\begin{figure}
\centering
\includegraphics[width=.45\textwidth]{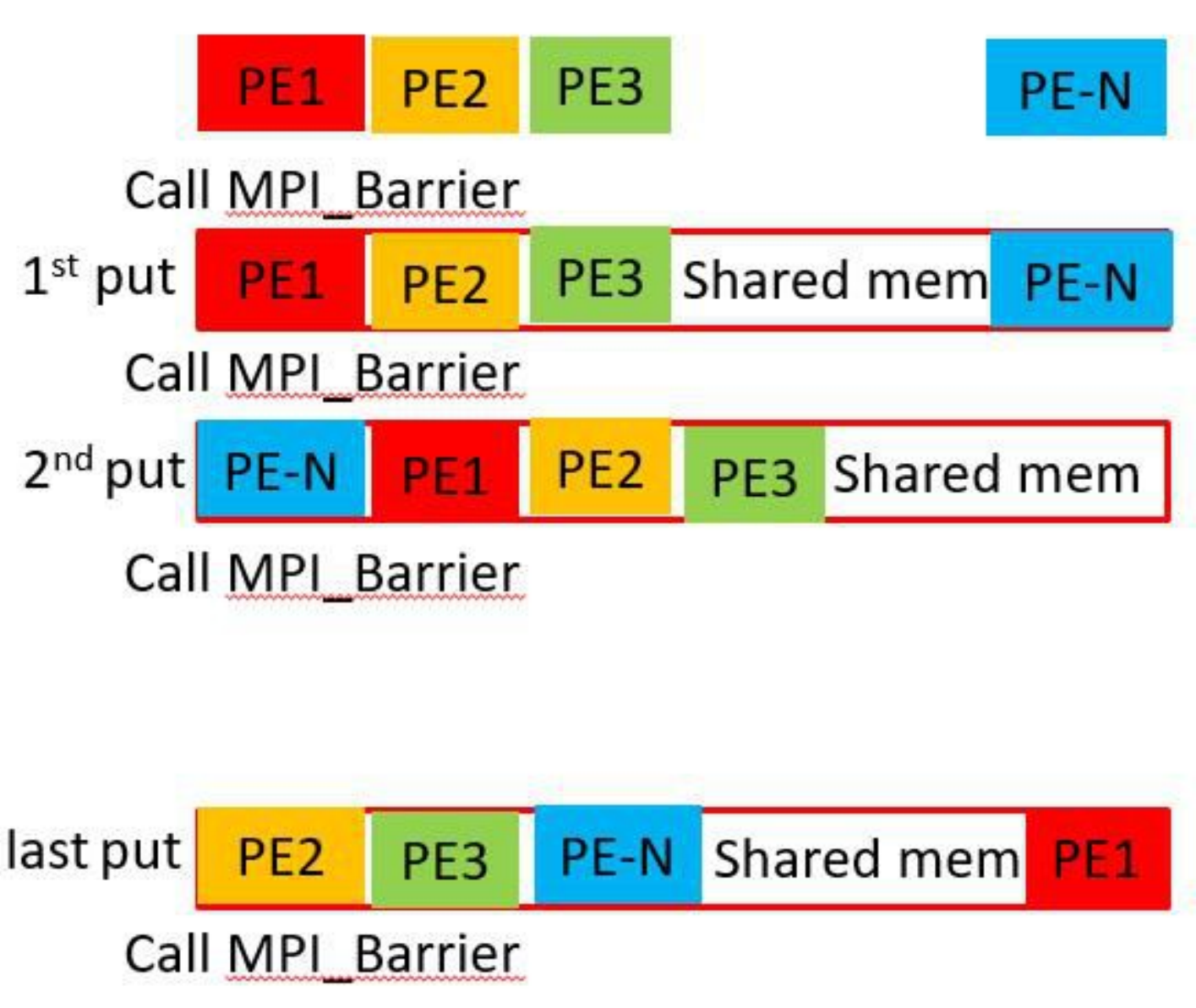}
\caption{\label{fig:shared_mem} The diagram of the marker projection to field variables on shared memory.}
\end{figure}

While markers are distributed among all processes and pushed forward in each process (particle decomposition), each process has the access to the whole set of the field variables (without domain decomposition). The field variables are stored in the shared memory on each computing node supported by the MPI-3 standard. This is motivated by the full $f$ simulations, for which the marker number is large and the cost on markers is the biggest part. In addition, the time step size ($\Delta t\sim T_{{\rm TAE}}/20$) allowed is significantly larger using mixed variables and the pullback scheme than the traditional pure $p_\|$ form, and a big potion of markers might migrate along magnetic field lines, which leads to significant computational cost if domain decomposition is otherwise adopted. The field equations, however, are solved using PETSc and the field solver is fully parallelized. While the atomic operations such as ``MPI\_ACCUMULATE'' on the shared memory are supported by MPI-3, the communication cost is still inefficient for marker-field projection and thus, the projection operation in TRIMEG-GKX is performed by binning the markers according to which part of memory they are to be written to, as shown in Fig. \ref{fig:shared_mem}.  Since the computational cost is mainly on the calculation of the basis function values on the marker location, the additional cost in determining whether it is the turn to put data on memory is negligible.

\section{Simulation setup and results}
\label{sec:results}
The toroidicity induced Alfv\'en eigenmode driven by energetic particles is simulated using the parameters defined by the ITPA group \cite{konies2018benchmark}. 
 The major radius $R_0=10 \; {\rm m}$, minor radius $a=1$ m, on-axis magnetic field $B_0=3$ T, the safety factor profile $q(r)=1.71+0.16r^2$. The electron density is constant with $n_{{\rm e}0}=2.0\times10^{19}\;\rm{m}^{-3}$, $T_{\rm{e}}=1$ keV. The ratio of the electron pressure to the magnetic pressure is $\beta_{\rm e}\approx 9\times 10^{-4}$. The Larmor radius of the thermal ion is $\rho_{\rm{ti}}=cm_{\rm{i}}v_{\rm{ti}}/(eB_{\rm{axis}})=0.00152$ m. The ratio between the adiabatic part ($\delta A_h/d_{\rm e}^2$) and the non-adiabatic part ($\nabla_\perp^2\delta A^{\rm{h}}$) in the left hand side of Amp\'ere's equation is $1/(d_{\rm e}^2k_\perp^2)\approx \beta_{\rm e}/(k_\perp\rho_{\rm{ti}})^2*(m_{\rm{i}}T_i/m_{\rm{e}}T_{\rm{e}})\approx1.622\times10^3$, where $k_\perp\approx nq/r=6\times1.75/0.5=21$.  This ITPA-TAE case is featured with a small electron skin depth ($d_{\rm e}\approx1.182\times10^{-3}$ m) and suffers from the ``cancellation problem'' if the pullback scheme is not adopted. 
 
 The EP density profile is given by 
\begin{align}
\label{eq:nEP1d}
	n_{\rm{EP}}(r)=n_{\rm{EP},0}c_3 \exp\left( -\frac{c_2}{c_1} \tanh\frac{r-c_0}{c_2}\right)\;\;, \\
	\frac{{\rm d}\ln n_{\rm{EP}}}{{\rm d}r} = -\frac{1}{c_1} \left[ 1-\tanh^2\frac{r-c_0}{c_2} \right]
\end{align}
where $n_{\rm{EP},0} =1.44131\times10^{17}\; \rm{m}^{-3}$, the subscript `$\rm{EP}$' indicates EPs (energetic particles), $c_0 = 0.491 23$, $c_1 =0.298 228$, $c_2 =0.198 739$, $c_3 =0.521 298$. The EP temperature is $400\; {\rm keV}$ for the base case.

\subsection{Numerical verification}
\begin{figure}
\centering
\includegraphics[width=.48\textwidth]{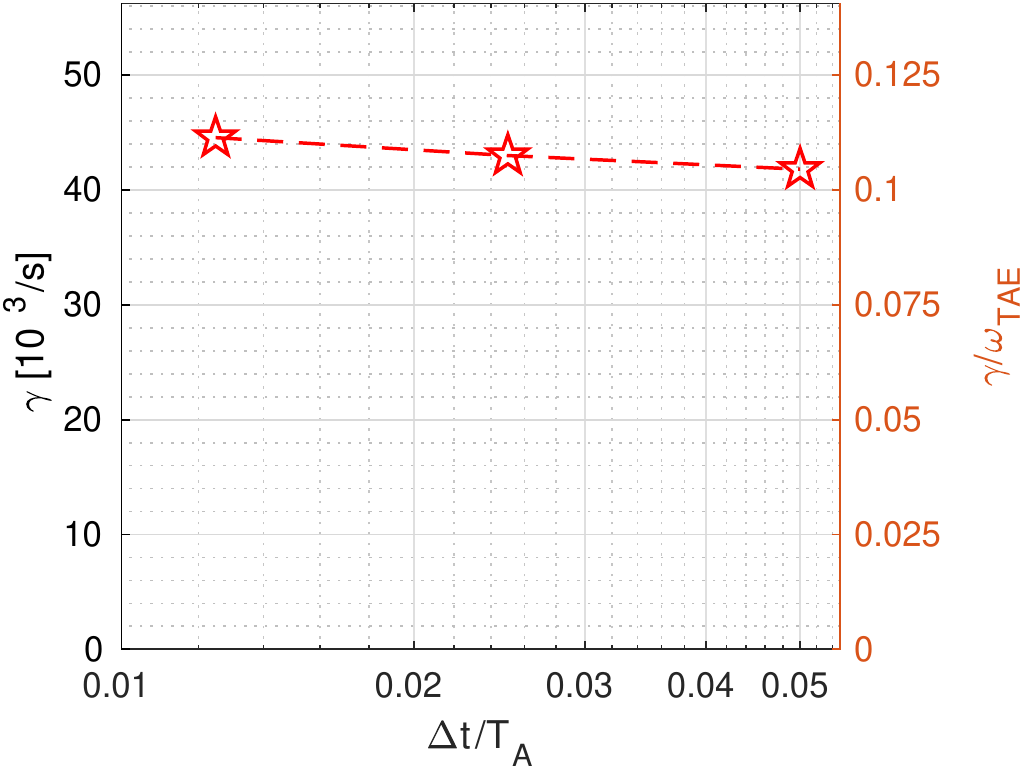}
\includegraphics[width=.48\textwidth]{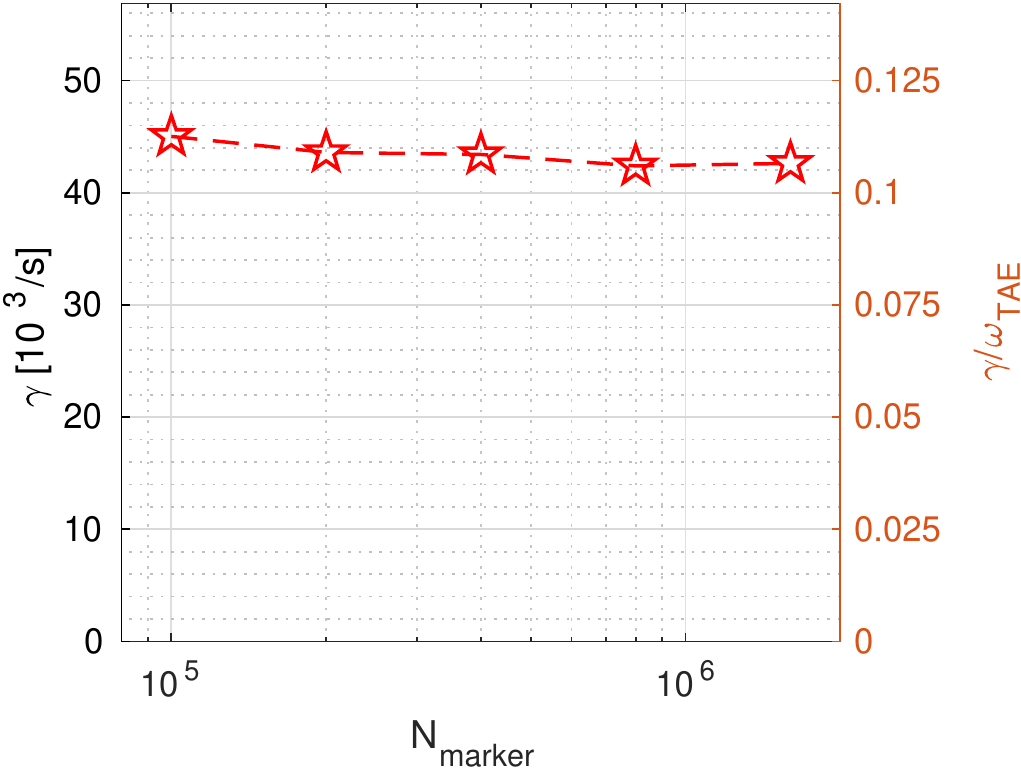}
\caption{\label{fig:scan4dtnptot}The growth rate for different values of the time-step size (left) and the marker number (right).}
\end{figure}

The base case of the EP driven TAE ($T_{\rm{EP}}=400$ keV) is used for the convergence studies. First, we need to identify the time step size and know its maximum acceptable value in the following studies. In this work, we simulate the $n=6$ TAE and only take $1/6$ of the torus in toroidal direction. Since using $8$ grids per wavelength is sufficient, we take $N_\phi=8$. Since the frequency is mainly determined by non-resonant particles while the growth rate is mainly determined by resonant EPs, the growth rate usually requires better resolution and we only show the convergence with respect to the growth rate. For typical grid size $(N_r,N_\theta,N_\phi)=(64,128,8)$, we choose the marker number $N_{\rm{marker}}=4\times10^5$ for each species. The growth rate for various time step size is shown in the left frame of Fig. \ref{fig:scan4dtnptot}. The growth rate starts to converge for $\Delta t/T_{\rm{A}}\le 0.05$, where $T_{\rm{A}}$ is the TAE period estimated at $r_{\rm c}=0.5$, i.e.,  $T_{\rm{A}}=4\pi q(r=r_{\rm c}) R_0/v_{\rm A}$, where $q(r=r_{\rm c})=q_{\rm c}=1.75$. For the convergence test related to the marker number, we choose $\Delta t/T_{\rm{A}}=0.05$ and the result is shown in the right frame of Fig. \ref{fig:scan4dtnptot}. Good convergence is achieved for $N_{{\rm marker}}>2\times10^5$. 

\begin{figure}
\centering
\includegraphics[width=.48\textwidth]{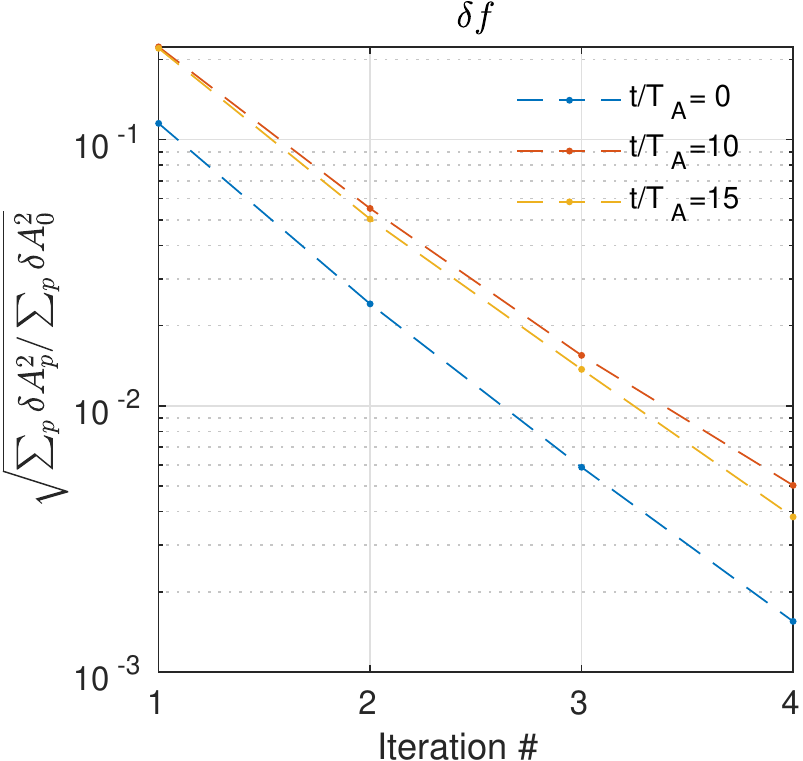}
\caption{\label{fig:convergence}The convergence of the iterative Amp\'ere solver.}
\end{figure}

The iterative Amp\'ere solver in Eq. \ref{eq:ampere_iterative} is crucial for the accurate calculation of $\delta A^{\rm{h}}$ and for the mitigation of the ``cancellation'' problem. 
The convergence of the iterative Amp\'ere solver of a typical nonlinear run is shown in Fig. \ref{fig:convergence}.
Good convergence is observed for the base case ($T_{\rm{EP}}=400$ keV). The EP driven TAE is excited and reaches the saturation after $t/T_{\rm{A}}\approx10$. The correction in $\delta A^{{\rm h}}$ is smaller as the number of iterations increases. At the initial state, the convergence is better than at later time since the marker distribution deviates away from Maxwell distribution due to the finite orbit width effect and mirror force, which leads to a larger discrepancy of the $\delta A^{\rm{h}}_0$ from the rigorous solution $\sum_{p=0,1,2\ldots}\delta A^{\rm{h}}_p$. Nevertheless, the convergence is good in the whole simulation and the correction to $\delta A^{\rm{h}}$ is suppressed to lower than $1\%$ in only 4 iterations. 

\subsection{Linear $\delta f$ simulations of EP driven TAEs}
\begin{figure}
\centering
\includegraphics[width=.48\textwidth]{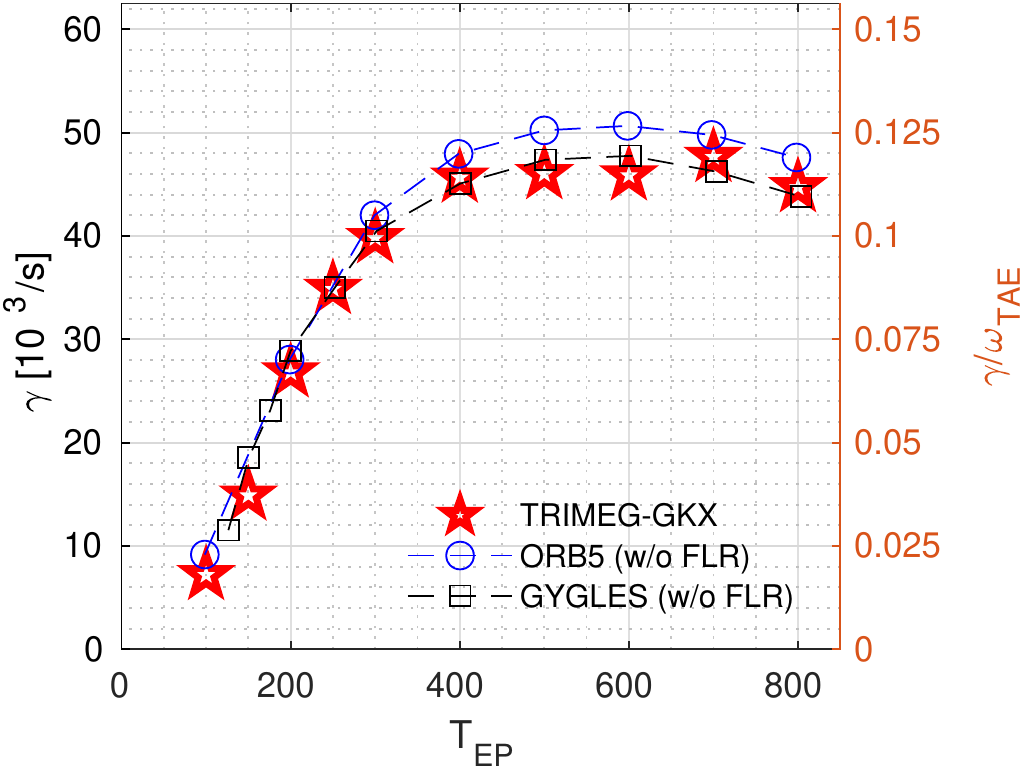}
\caption{\label{fig:gamma_scan} The growth rate of EP driven TAEs for different values of EP temperature for models without finite Larmor radius (FLR) effect. }
\end{figure}

The EP driven TAE is simulated with various EP temperature for the benchmark with other codes in previous work \cite{konies2018benchmark}.
To make the model in TRIMEG-GKX as similar as in other codes, the $\delta f$ scheme is applied to thermal ions, energetic particles and electrons. Since we focus on the linear solution, the linear gyrokinetic equations are solved with markers pushed along the unperturbed trajectory but the marker weight evolving along time.  
We only compare with ORB5 and GYGLES results since the models are more similar to ours. 
The growth rate of EP driven TAE is shown in Fig. \ref{fig:gamma_scan}. Good agreement is achieved, for the model  without finite Larmor radius effect.  

\begin{figure}
\centering
\includegraphics[width=.48\textwidth]{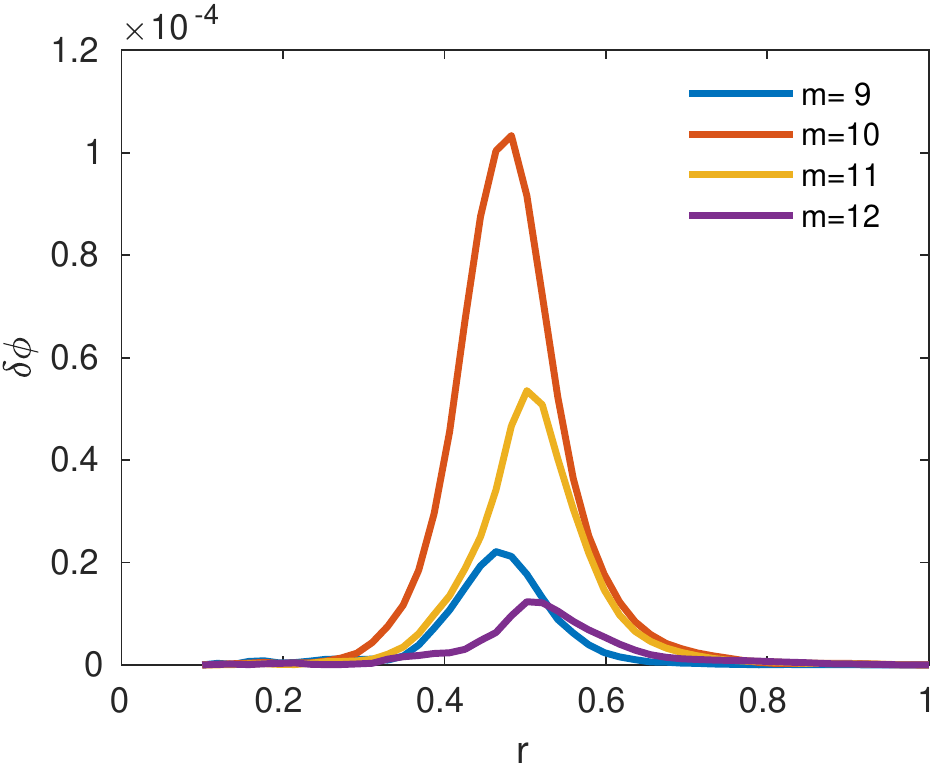}
\caption{\label{fig:frmphi} The radial structure of poloidal harmonics for EP driven TAE,  with EP temperature $T_{\rm{EP}}=400$ keV. The four dominant harmonics are plotted ($m=9,10,11,12$).}
\end{figure}

The radial mode structures of the poloidal harmonics for $T_{\rm{EP}}=400$ keV are shown in Fig. \ref{fig:frmphi}. The mode width is comparable to that from ORB5/GYGLES \cite{konies2018benchmark}. The magnitude of the $m=10$ harmonic is significantly higher than that of the $m=11$ harmonic, which is also consistent with the results of GYGLES/ORB5 as shown in the previous work \cite{konies2018benchmark}.

\subsection{Nonlinear $\delta f$ and full $f$ EP simulations}
Nonlinear simulations are performed using the $\delta f$ scheme and the mixed-full-$f$-EPs-$\delta f$-electrons/thermal-ions scheme separately. Note that in the mixed full $f$-$\delta f$ scheme, the full $f$ scheme is only applied to EPs but electron and thermal ions are always treated using the $\delta f$ scheme. To estimate the minimum marker number for proper treatment of TAE, it is noticed that $\delta\bar{n}\sim\delta\bar{\phi}(k_\perp^2\rho_{\rm{ti}}^2)\sim10^{-3}\delta\bar{\phi}$,
$\delta\bar{j}\sim\delta\bar{A}_\|(\beta m_{\rm{i}}/m_{\rm{e}})/\rho_{\rm{ti}}^2\sim6.2\times10^{-4}\delta\bar{A}$ where $\delta\bar{n}$ and $\delta\bar{j}$ are normalized using electron equilibrium density and electron thermal velocity. At the saturation level, $\delta\bar{\phi}\sim1$, $\delta\bar{A}_\|\sim1$. If the full $f$ is adopted for all species, the noise level in density and current should be much lower than $\delta \bar{n}$ and $\delta\bar{j}$ which yields that the marker number per degree of freedom (the product of the cell number and the Fourier mode number) $N_{\text{mark, DOF}}\gg 10^6$ for density and $N_{\text{mark, DOF}}\gg 2.6\times10^6$ for the current.
In order to reduce the marker number but keep the capability of treating EPs using the full $f$ scheme, in this work we adopt the $\delta f$ scheme for electrons and thermal ions. Thus the main noise is from the full $f$ EPs. Since the EP density is less than $1\%$ of the electron density, the criteria for the marker number is relaxed to that the marker number per cell or per Fourier mode $N_{\text{mark, DOF}}\gg 10^2$ for density and $N_{\text{mark, DOF}}\gg 2.6\times10^2\sqrt{T_{\rm{EP}} m_{\rm{e}}/(T_{\rm{e}} m_{\rm{EP}})}\sim0.86\times10^2$ for the current.

\begin{figure}
\centering
\includegraphics[width=.33\textwidth]{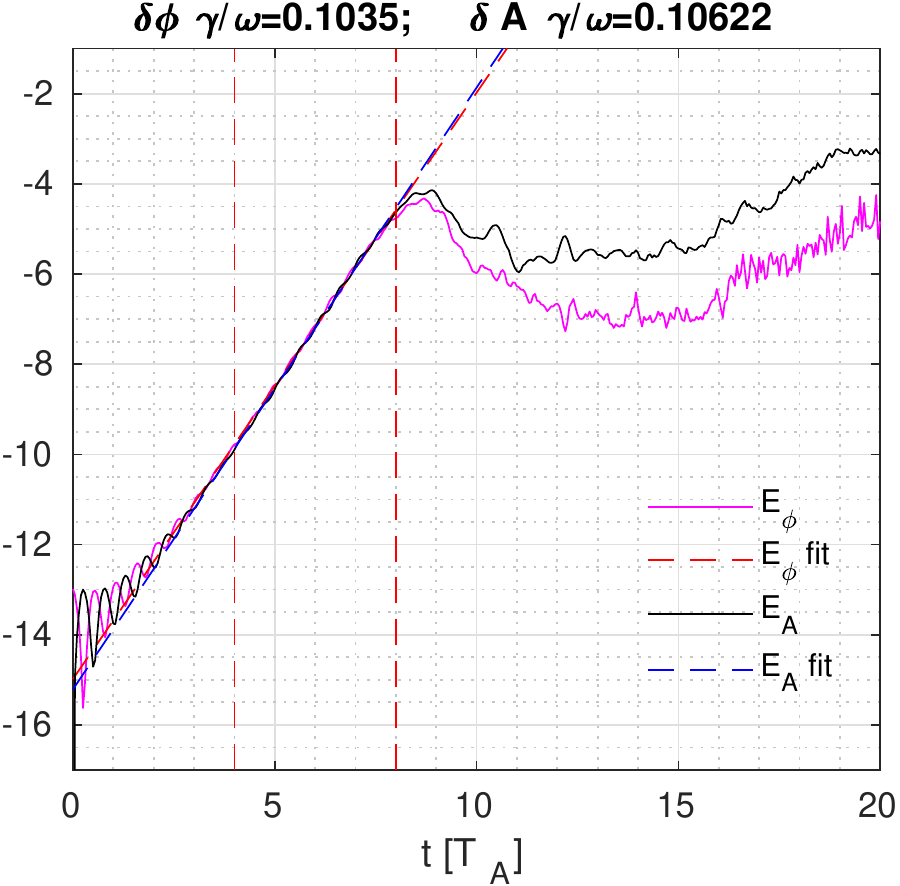}
\includegraphics[width=.33\textwidth]{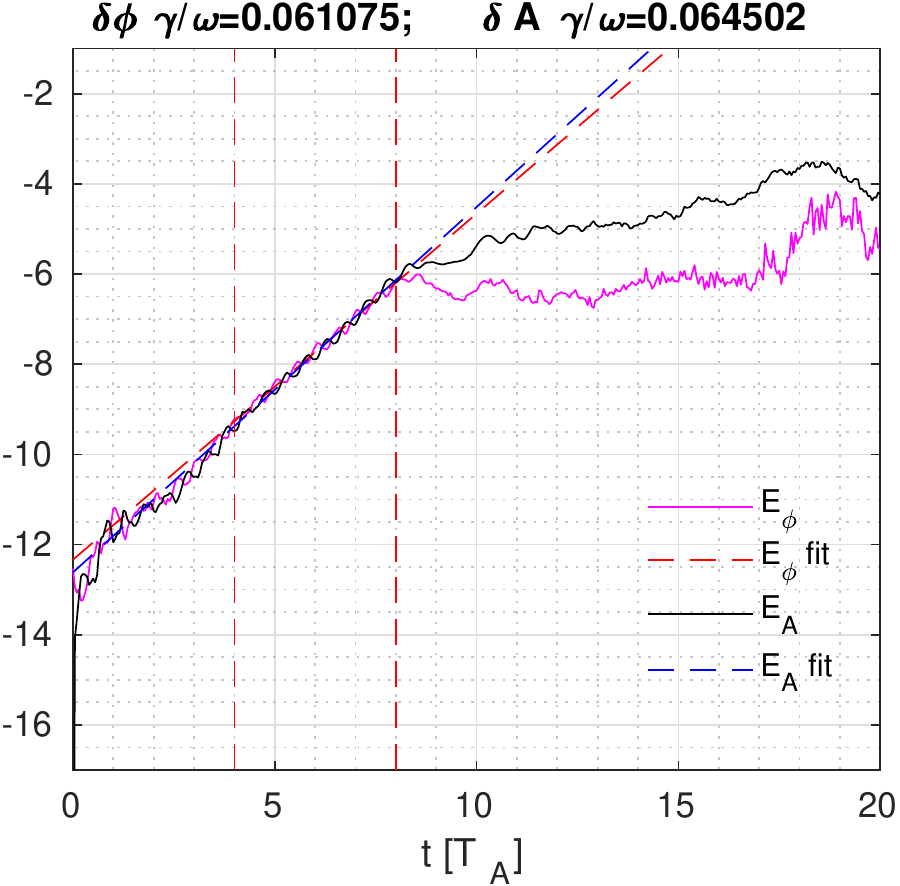}
\includegraphics[width=.33\textwidth]{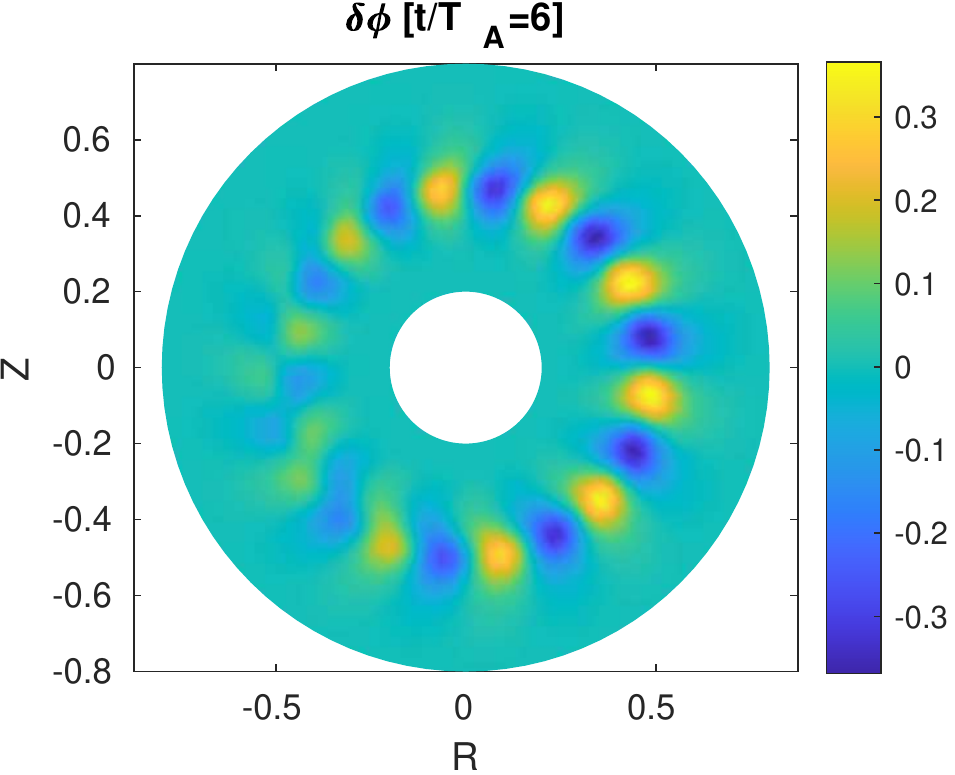}
\includegraphics[width=.33\textwidth]{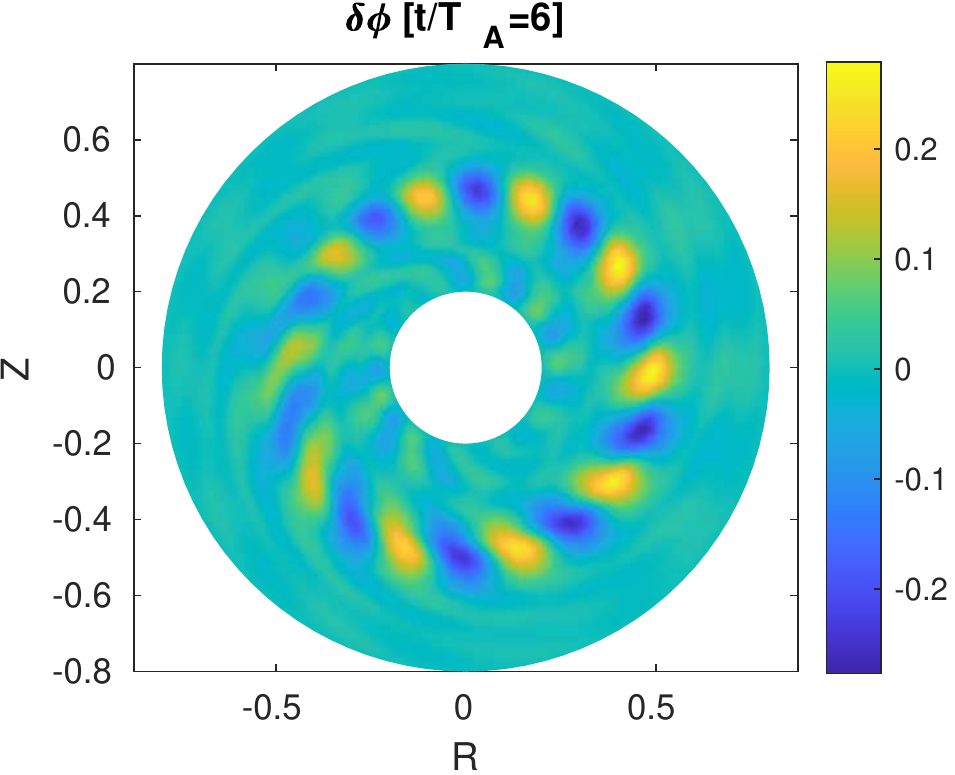}
\caption{\label{fig:grwothNL}The time evolution of $\delta f$ (top left) and full $f$ (top right) simulations. The corresponding 2D mode structures in the end of the linear stage are shown at the bottom. }
\end{figure}

The time evolution of the field energy is shown in Fig. \ref{fig:grwothNL}. 
The noise level is controlled with the marker numbers $N_{\rm{e}}=2.5\times10^5$, $N_{\rm{i}}=2.5\times10^5$, $N_{{\rm{EP}}}=16\times10^6$ in the full $f$ simulations. In the $\delta f$ simulation, $N_{\rm{e}}=2.5\times10^5$, $N_{\rm{i}}=2.5\times10^5$, $N_{{\rm{EP}}}=2.5\times10^5$. The main computational consumption is for the operations related to markers, namely, the interpolation of field at the marker location, the calculation of the density and the current using markers, and the calculation of the marker trajectories. Consequently, the full $f$ case is about 20 times more expensive than the $\delta f$ case, which is consistent with the marker numbers $16.5\times10^6$ in the full $f$ case and $0.75\times10^6$ in the $\delta f$ case.  Clear linear growth stage and mode saturation are observed. The mode structure of the linear stage is in good agreement with the $\delta f$ simulations.
As far as we know, this mixed scheme with full $f$ EPs and $\delta f$ thermal ions and electrons has not been reported before. In our studies, this method has been validated to be practical and suitable for the studying EP related physics. As we estimate, the simulation time is reduced to $(n_{\rm{EP}}/n_{\rm{e}})^2\sim10^{-4}$ of that if the full $f$ scheme is applied to electrons and ions.

The linear growth rate and saturation level of the full $f$ simulation are relatively smaller than those of the $\delta f$ simulation, due to the weaker driving strength caused by the EP profile relaxation.
Note that in both the full $f$ and the $\delta f$ simulations, the same particle refilling scheme has been applied, namely, the lost particles are refilled at the poloidal location $\theta_{p,\mathrm{refill}}=-\theta_{p,\mathrm{loss}}$, and at the toroidal location $\phi_{p,\mathrm{refill}}=\phi_{p,\mathrm{loss}}-2q(r_{p,\mathrm{loss}})\theta_{p,\mathrm{loss}}$, where $r_{p,\mathrm{loss}}$ indicates the radial location of the lost particles. 
The EP profile relaxation is from the marker loading process. The local Maxwellian distribution is not a steady state solution in tokamak geometry, namely, the distribution function relaxes in a few particle transit period ($\sim 2\pi q R_0/v_{\|}$). The EP profile relaxation in the full $f$ scheme should be treated rigorously by adopting the EP distribution in constant of motion space, in order to eliminate artificial ingredient for benchmark with different codes and for interpreting experimental observations.

\subsection{Full $f$ simulation using canonical Maxwellian distribution}
\begin{figure}
\centering
\includegraphics[width=.88\textwidth]{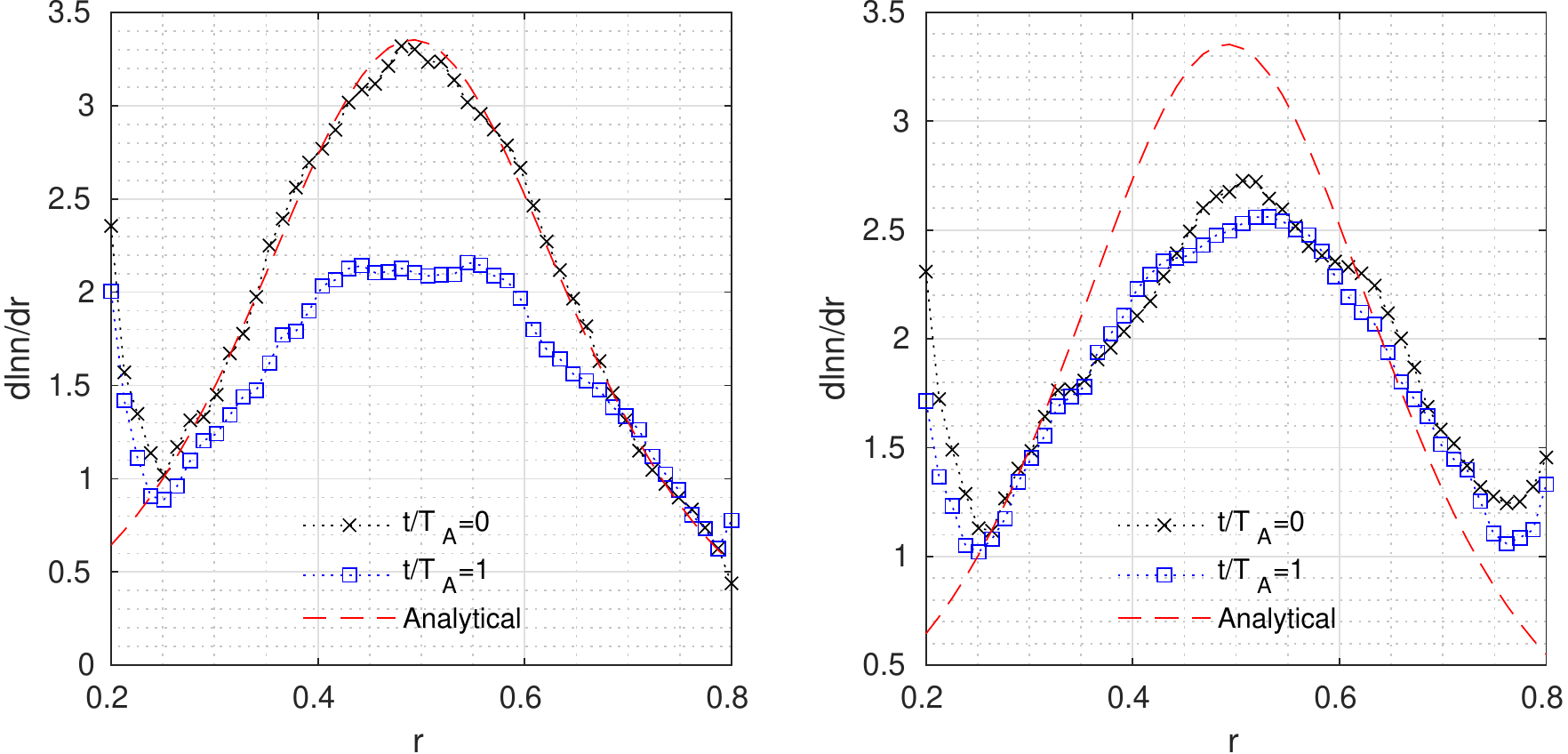}
\caption{\label{fig:EPrelax} The change of marker distribution from $t/T_{\rm{A}}=0$ to $t/T_{\rm{A}}=1$ for the local Maxwell distribution (left) and canonical Maxwell distribution (right). The marker relaxation is significant for the local Maxwell distribution but negligible for canonical Maxwell distribution. For both case, $(c_0,c_1,c_2,c_3)=(0.49123,0.298228,0.198739,0.521298)$.  }
\end{figure}

While the full $f$ simulations using a local Maxwellian distribution is a well defined problem and can be used as a benchmark case for full $f$ studies, it is important to adopt a kinetic equilibrium based on the constants of motion (canonical Maxwellian distribution), and to identify the differences between simulations using the canonical Maxwellian distribution and local one. The EP profile relaxation is shown in Fig. \ref{fig:EPrelax} for cases using the local Maxwellian distribution and the canonical Maxwellian distribution. The logarithmic EP gradient decreases by $\sim40\%$ in $1\sim2$ EP transit periods for the local Maxwellian distribution (left frame) but stays all most the same for the canonical distribution (right frame)  ($T_{\rm{trans,EP}}/T_{\rm{A}}\sim 1$ for the base case with $T_{\rm{EP}}=400\;\rm{keV}$). 
However, the canonical distribution should be adjusted to match the analytical radial distribution since the EP distribution function using the shifted canonical toroidal momentum $\psi_{\rm{can}}$ describes the distribution of the EP orbit center. 

\begin{figure}
\centering
\includegraphics[width=.45\textwidth]{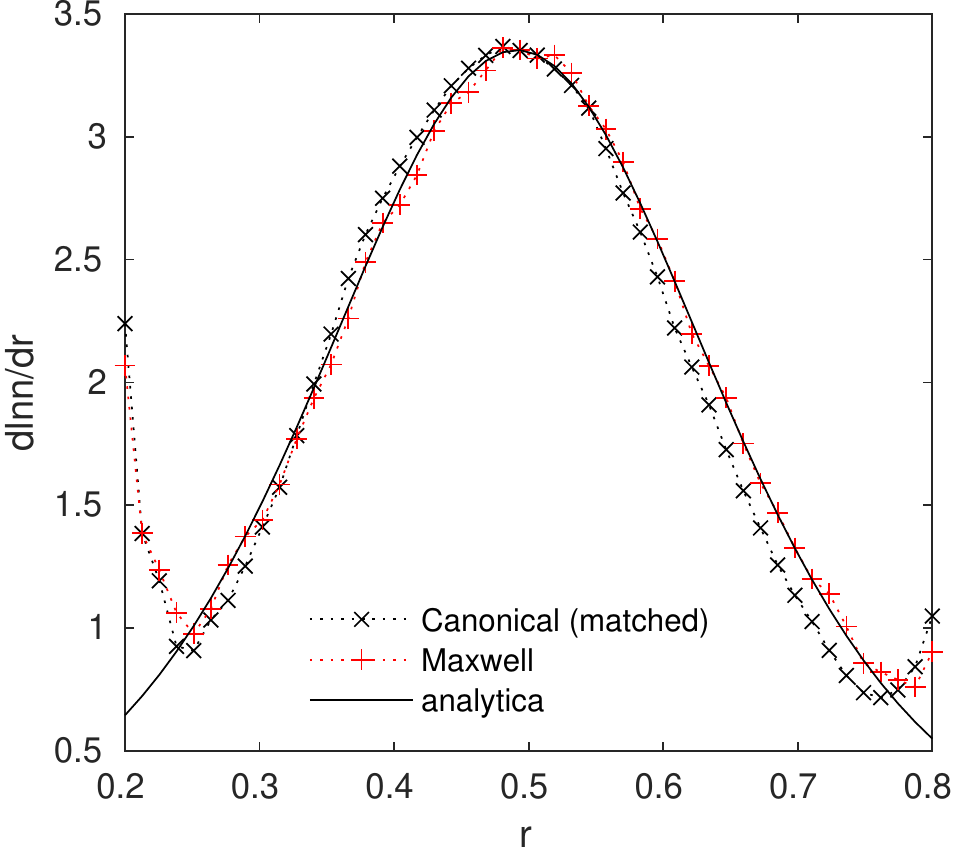}
\includegraphics[width=.45\textwidth]{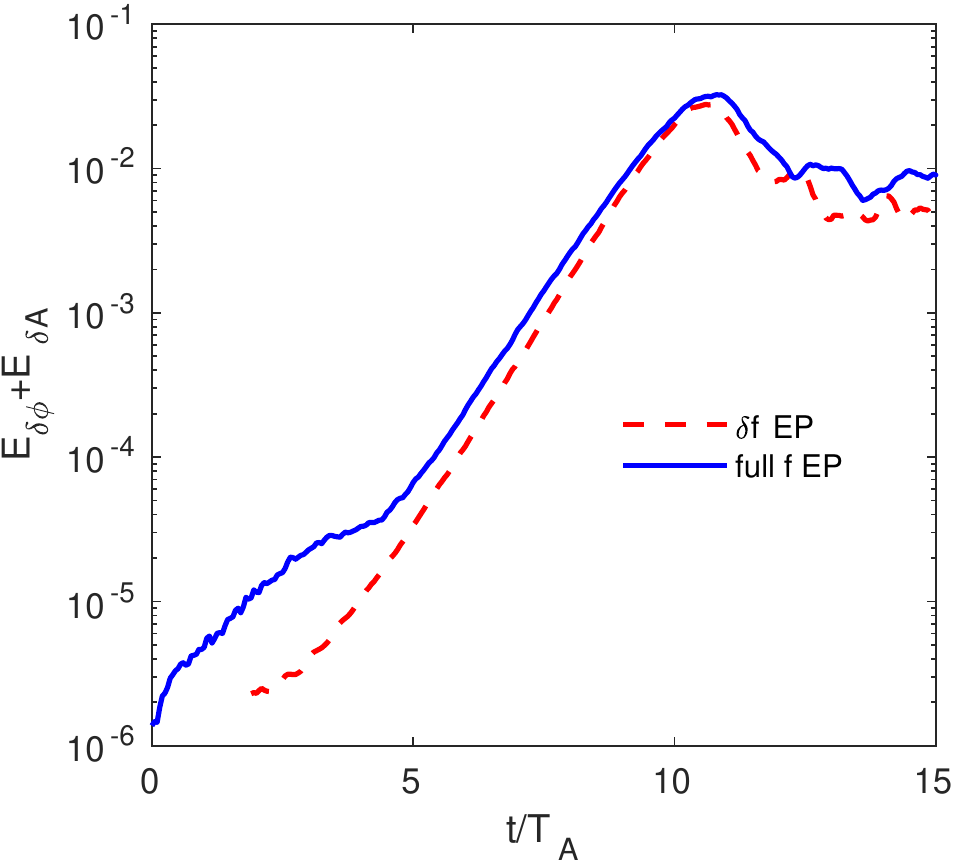}
\caption{\label{fig:matchCanonicalMaxwell} Left: the matched canonical Maxwell distribution with $(c_0,c_1,c_2,c_3)=(0.46623,0.17042,0.11357,0.521298)$ (black crosses); the local Maxwell distribution with $(c_0,c_1,c_2,c_3)=(0.49123,0.298228,0.198739,0.521298)$ from marker (red plus) and from analytical results (black line).
Right: time evolution of total field energy from $\delta f$ simulation (red dashed line) using local Maxwell EP distribution and from a full $f$ simulation (blue line) using a canonical Maxwell EP distribution. }
\end{figure}

While the EP guiding center distribution is given by Eq. \ref{eq:nEP1d}, the conversion between the EP guiding center and the EP orbit center can be obtained theoretically by applying the push-forward and pull-back transform between the guiding center and the orbit center. In this work, the matching procedure can be done numerically. Several cases with canonical EP distribution featured by different values of $c_0,c_1,c_2,c_3$  are run and adapted so that the density profile of the guiding center matches that in Eq. \ref{eq:nEP1d}. A good match is obtained as we choose $(c_0,c_1,c_2,c_3)_{\rm{can}}=(0.46623,0.17042,0.11357,0.521298)$ as the coefficients in the canonical EP distribution in Eq. \ref{eq:fcanEP}
\begin{eqnarray}
    n(\psi_{\rm{can}})=n(r_{\rm{can}})=n_{\rm{{EP,0}}}c_{3,\rm{can}} \exp\left( -\frac{c_{2,\rm{can}}}{c_{1,\rm{can}}} \tanh\frac{r-c_{0,\rm{can}}}{c_{2,\rm{can}}}\right) \;\;.
\end{eqnarray}
For the full $f$ case, the  marker numbers $N_{\rm{e}}=2.5\times10^5$, $N_{\rm{i}}=2.5\times10^5$, $N_{{\rm{EP}}}=64\times10^6$.
The guiding center profile of this matched case is shown in the left frame of Fig. \ref{fig:matchCanonicalMaxwell}. It is very close to the nominal profile described by the ITPA-TAE case in Eq. \ref{eq:nEP1d}. More precise matching can be achieved in principle by representing the EP profile using finite element method in constant of motion space but is beyond the scope of this work. 
In addition, the artificial relaxation is avoided. In the right frame, the time evolution of the full $f$ scheme agrees with that of the $\delta f$ scheme. The difference in growth rate is still observed, but is only by $\sim8\%$, since the detailed structure of the distributions in full $f$ and $\delta f$ and the related wave-particle interaction can be different. Nevertheless, the application of the canonical EP distribution is shown to be a practical way to avoid the large EP profile relaxation and to match the EP density profile in the full $f$ simulations. For this ITPA-TAE case, the EP density perturbation is small $\delta n/n_0<1\%$ and the $\delta f$ scheme is still applicable; thus it is suitable to compare the full $f$ and the $\delta f$ schemes. For longer time scale simulations and intermittent and transient plasmas, the EP profile can vary more significantly and the full $f$ EP scheme provides a natural way due to its capability of describing arbitrary distribution and its time evolution in phase space but the  $\delta f$ becomes less powerful due to the larger $\delta f/f_0$ and the consequent enhanced noise level. More simulations with more significant time evolution of the EP distribution will be our future work. 

\section{Conclusions}
In this work, the full $f$ and $\delta f$ gyrokinetic particle models have been formulated and implemented on the same footing and applied to the simulation of toroidicity induced Alfv\'en eigenmode driven by energetic particles. 
The mixed full $f$-$\delta f$ scheme has been proposed for the studies of Alfv\'en waves and energetic particle physics. 
The mixed variables have been adopted in the formulation and the pullback scheme is implemented in TRIMEMG-GKX code. 
Excellent performance has been demonstrated for the simulation of the electromagnetic problems and the cancellation problem is mitigated even for the ITPA-TAE case for which the electron skin depth is small. 
Good agreement with previous results from other codes is observed in terms of linear growth rate of EP driven TAEs and mode structures. 

Important physics properties of the energetic particles have been demonstrated by the full $f$ (EP) simulations. Due to the high energy of EPs, the finite orbit width is large and the local Maxwell distribution is not a good approximation of the steady state EP distribution. 
Using the Maxwell EP distribution in the full $f$ simulations, the EP distribution deviates from the initial one in a few transit periods, leading to a EP profile relaxation featured by weakened EP density gradient.
Consequently, the full $f$ EP simulations gives a linear growth rate smaller than the $\delta f$ simulations by $40\%$ for $T_{\rm{EP}}=400$ keV using the ITPA-TAE parameters, which suggests that the EP profile relaxation in the full $f$ scheme should be treated properly. The shifted canonical toroidal moment, the particle energy and the magnetic momentum are adopted in this work as the constants of motion of EPs and the EP distribution defined on these coordinates gives a more rigorous description of the EP steady state, without suffering from artificial density relaxation. In addition, a matching scheme has been introduced for demonstrating the pullback transform from the shifted canonical toroidal momentum coordinate to the EP guiding center coordinate, showing the essence of the EP properties due to the finite orbit width effect.  The mixed-full-$f$-$\delta f$ scheme developed in this work makes the kinetic simulations of the background electrons and thermal ions computationally economical and meanwhile brings in a flexible description for EPs which allows a large variation of EP profiles and distributions in velocity space. It provides a powerful tool for kinetic studies using realistic experimental EP distributions related to intermittent and transient plasma activities. 

\begin{acknowledgments}
The fruitful discussions with A. Mishchenko, F. Zonca, M. Rampp, and J. Chen and inputs from JOREK group, ORB5 group and EUTERPE group are appreciated by Z.X. Lu. The simulations in this work are run on MPCDF. 
This work has been carried out within the framework of the EUROfusion Consortium, funded
by the European Union via the Euratom Research and Training Programme (Grant Agreement No
101052200 — EUROfusion). Views and opinions expressed are however those of the author(s)
only and do not necessarily reflect those of the European Union or the European Commission.
Neither the European Union nor the European Commission can be held responsible for them.
\end{acknowledgments}

\appendix
\section{Ad-hoc equilibrium}
For the tokamak geometry, the coordinates $(r,\phi,\theta)$ are adopted and the magnetic field is represented as ${\bf B}=\nabla\psi\times\nabla\phi+F\nabla\phi$ and ${\bf b}={\bf B}/B$ is the unit vector along the field lines. An ad-hoc equilibrium has been adopted, featured with concentric circular magnetic flux surfaces and constant $F$. The poloidal flux
$$\psi(r)=\frac{B_0}{2\bar{q}_2}\ln{(1+\frac{\bar{q}_2}{\bar{q}_0}r^2)}, $$
where $\bar{q}(r)=\bar{q}_0+\bar{q}_2r^2$, $\bar{q}_2=\bar{q}_{\rm{edge}}-\bar{q}_0$. The function of safety factor
$$q(r)=\frac{\bar{q}}{\sqrt{1-(r/R_0)^2}} .$$
The curl of the magnetic field direction

\begin{align}
     &(\nabla\times{\bf b}) \cdot \hat{{\bf r}}=\frac{R_0\sin\theta}{R\sqrt{r^2/\bar{q}^2+R_0^2}}   \\
     &(\nabla\times{\bf b}) \cdot \hat{\boldsymbol{\theta}}=
     \frac{R_0}{R} \left[ \frac{\cos\theta}{\sqrt{r^2/\bar{q}^2+R_0^2}} -  \frac{Rr(\bar{q}_0-\bar{q}_2r^2)}{\bar{q}^3(r^2/\bar{q}^2+R_0^2)^{3/2}} \right]    \\
     &(\nabla\times{\bf b})\cdot\hat{\boldsymbol{\phi}}= -\frac{r^2+2R_0^2\bar{q}_0\bar{q}}{\bar{q}^3(r^2/\bar{q}^2+R_0^2)^{3/2}} 
\end{align}

\section{Guiding center equations of motion in ad-hoc equilibrium}
The equilibrium part of the motion is as follows,
\begin{align}
    \frac{{\rm d} r_0}{{\rm d}t} &= b^*_r v_\| + C_{{\rm d}} \frac{F}{r} \partial_\theta B\;\;, 
    \\
    \frac{{\rm d}\theta_0}{{\rm d}t} &= \frac{b^*_\theta}{r}v_\| - C_{{\rm d}} \frac{F}{r} \partial_r B \;\;, 
    \\
    \frac{{\rm d}\phi_0}{{\rm d}t} &= \frac{b^*_\phi}{R} v_\| +C_{{\rm d}} \frac{\partial_r\psi}{R} \partial_r B \;\;,
    \\
    &C_{{\rm d}}=\rho_{\rm{N}}\frac{B_{\rm{axis}}}{RB^2 B^*}\frac{M_s}{\bar{e}_s}\mu B \;\;,
    \\
    \frac{{\rm d}v_{\|,0}}{{\rm d}t} &= -\frac{\mu b^*_\theta}{r}\partial_\theta B \;\;.
\end{align}

The perturbed part of the equations of motion is
\begin{align}
    \frac{{\rm d}\delta r}{{\rm d}t} 
    & = C_{\rm{E}} (b_\phi\partial_\theta \delta G -b_\theta \partial_\phi\delta G) 
    \;\;,
    \\
    \frac{{\rm d}\delta \theta}{{\rm d}t} 
    &= - \frac{C_{\rm{E}}}{r} b_\phi\partial_r \delta G 
     \;\;,
    \\
    \frac{{\rm d}\delta \phi}{{\rm d}t} 
    &= \frac{C_{\rm{E}}}{R} b_\theta\partial_r \delta G  \;\;,
    \\
    &\delta G=\delta\phi-v_\|\delta A_\| \;\;
    \\
    \frac{{\rm d}\delta v_\|}{{\rm d}t} 
    &=  C_{\|1}(
     \frac{b^*_r}{r}      v_\|\partial_r      \delta A^{\rm{h}} 
    +\frac{b^*_\theta}{r} v_\|\partial_\theta \delta A^{\rm{h}}  
    +\frac{b^*_\theta}{R} v_\|\partial_\phi \delta A^{\rm{h}} 
    )
    \\
    &+C_{\|2} (
     \frac{b_\phi}{r}\partial_\theta B\partial_r\delta A^{\rm{s}}
    -\frac{b_\phi}{r}\partial_r B\partial_\theta\delta A^{\rm{s}}
    +\frac{b_\theta}{R}\partial_r B\partial_\phi\delta A^{\rm{s}}
    )
    \\
    &C_{\rm{E}} = \rho_{\rm{N}} \frac{B_{\rm{axis}}}{B^*}\;\;, \;\;
    C_{\|1} = \frac{\bar{e}_s}{M_s} \;\;,\;\;
    C_{\|2} = -\rho_{\rm{N}}\mu \frac{B_{\rm{axis}}}{B^*} \;\;.
\end{align}

\providecommand{\newblock}{}

\end{document}